\documentclass[fleqn,usenatbib]{mnras}
\usepackage{newtxtext,newtxmath}
\usepackage[T1]{fontenc}
\usepackage{ae,aecompl}

\usepackage{graphicx}	% Including figure files
\usepackage{amsmath}	% Advanced maths commands
\usepackage{amssymb}	% Extra maths symbols
\usepackage{xcolor}
\newcommand{\Angstrom}{\textup{\AA}}
\newcommand{\Fxuv}{F$_{\rm xuv}$}
\title[Effect of magnetic cycle on atmospheric escape]{Influence of the Sun-like magnetic cycle on exoplanetary atmospheric escape}

\author[Hazra, Vidotto \& Villarreal D'Angelo]{
Gopal Hazra,$^{1}$ \thanks{E-mail: gopal.hazra@tcd.ie}
Aline A. Vidotto$^{1}$
and Carolina Villarreal D'Angelo$^{2,1}$
\\
$^{1}$ School of Physics, Trinity College Dublin, the University of Dublin, College Green, Dublin, D-2, Ireland \\ % D02 PN40,
$^{2}$Observatorio Astron{\'o}mico de C{\'o}rdoba (OAC-UNC), Universidad Nacional de C{\'o}rdoba. Laprida 854, C{\'o}rdoba-Argentina%Postal Code, Country\\
}

\date{Accepted XXX. Received YYY; in original form ZZZ}

\pubyear{2020}

\begin{document}
\label{firstpage}
\pagerange{\pageref{firstpage}--\pageref{lastpage}}
\maketitle

\begin{abstract}
Stellar high-energy radiation (X-ray and extreme ultraviolet, XUV) drives atmospheric escape in close-in exoplanets. Given that stellar irradiation depends on the stellar magnetism and that stars have magnetic cycles, we investigate how cycles affect the evolution of exoplanetary atmospheric escape. Firstly, we consider a hypothetical HD209458b-like planet orbiting the Sun. For that, we implement the observed solar XUV radiation available over one and a half solar cycles in a 1D hydrodynamic escape model of HD209458b. We find that atmospheric escape rates show a cyclic variation (from 7.6 to 18.5 $\times$ 10$^{10}$ g~s$^{-1}$), almost proportional to the incident stellar radiation. To compare this with observations, we compute spectroscopic transits in two hydrogen lines. We find non-detectable cyclic variations in Ly$\alpha$ transits. Given the temperature sensitiveness of the H$\alpha$ line, its equivalent width has an amplitude of 1.9 m\AA\ variation over the cycle, which could be detectable in exoplanets such as HD209458b. We demonstrate that the XUV flux is linearly proportional to the magnetic flux during the solar cycle. Secondly, we apply this relation to derive the cyclic evolution of the XUV flux of HD189733 using the star’s available magnetic flux observations from Zeeman Doppler Imaging over nearly a decade. The XUV fluxes are then used to model escape in HD189733b, which shows escape rate varying from 2.8 to 6.5 $\times$ 10$^{10}$ g~s$^{-1}$. Like in the HD209458b case, this introduces variations in Ly$\alpha$ and H$\alpha$ transits, with H$\alpha$ variations more likely to be observable. Finally, we show that a strong stellar flare would enhance significantly Ly$\alpha$ and H$\alpha$ transit depths.
\end{abstract}


%%%%%%%%%%%%%%%%%%%%%%%%%%%%%%%%%%%%%%%%%%%%%%%%%%

%%%%%%%%%%%%%%%%% BODY OF PAPER %%%%%%%%%%%%%%%%%%

\section{Introduction}
%observational background

Hot Jupiters, a class of gaseous planets of mass comparable to the Jupiter  (1--10 M$_{\rm Jup}$), have been observed to go through strong atmospheric loss \citep[e.g.,][]{Vidal-Madjar2003, Linsky10, 2010ApJ...714L.222F, Ehrenreich2015}. Because these planets orbit very close to their host stars ($\lesssim$ 0.1 au) and are always exposed to intense radiation from their host stars, it is believed that photoionisation drives escape \citep[e.g.,][]{2003ApJ...598L.121L,2004A&A...418L...1L}. Stellar radiation ionises the planetary material and helps it to escape the planetary gravity in the form of a planetary wind. The detection of atomic hydrogen beyond the Roche distance first established that atmospheric material was escaping from  the hot Jupiter HD209458b \citep{Vidal-Madjar2003}. The estimated escape rate for this planet is $\dot{M}$ = $10^{10}-10^{11}$ g s$^{-1}$ \citep{Vidal-Madjar2003, Ehrenreich2008, Lampon2020}. Many subsequent observations \citep{Vidal-Madjar04, Linsky10,Cubillos2020} also showed the presence of atomic oxygen, ionized carbon, silicon, magnesium and ionized iron at  very high altitudes from the planet, adding more evidence that not only hydrogen but other heavier materials are also undergoing escape. Recently, transit spectroscopic observations of  other targets (e.g., HD189733b, Wasp-12b, HD189833b and Kelt-9b) found that atmospheric escape via photoevaporation is likely a general phenomenon in  hot Jupiters that could contribute to their evolution \citep{2015ApJ...815L..12J, 2018A&A...612A..25K, 2019A&A...632A..65K, Allan2019}. Atmospheric escape has been observed in warm Neptunes as well. A transit depth of 56.3 $\pm$ 3.5\% in the Ly$\alpha$ line (well above the optical transit depth of 0.69\%) was observed in the warm Neptune GJ436b, with estimated escape rates in the range of about $10^8 - 10^9$ g/s \citep{2014ApJ...786..132K,Ehrenreich2015,2017A&A...605L...7L}. All this shows that atmospheric escape is important for hot, hydrogen-rich gas giants, in particular for planets in the ‘hot Jupiter/warm Neptune' category.

Thanks to monitoring campaigns, it has been possible to verify that some planets show a significant amount of temporal variation in their  Ly$\alpha$ transits, as is the case of the hot Jupiter HD189733b. This transit variability has been interpreted as due to a change in the physical conditions of the evaporating atmosphere  \citep{Lecavelier12}. Temporal variation was also observed in the H$\alpha$ transit of HD189733b \citep{Barnes16}, confirming once more that evaporating atmospheres undergo temporal evolution. One likely explanation is that these temporal changes are caused by stellar variability, i.e., time-variations in the properties of host stars, which can take place in the form of, e.g., cyclic magnetic activity, strong flares or  variation in  stellar winds or total stellar output radiation \citep[e.g.,][]{Baliunas95,Tokumaru2010,BoroSaikia18,Hazra19}.

Hydrodynamic escape models of planetary atmospheres heated by photoionisation are able to obtain escape rates comparable to those derived from observations \citep{Murray-clay2009,Tian05,Koskinen13, Debrecht2020}. Apart from the stellar radiation,  close-in planets interact with the magnetized winds and magnetic fields of their host stars \citep{Lanza13, Carroll-Nellenback2017, Villarreal2018,Esquivel2019,McCann19}, which can change the observable signatures of  transiting exoplanetary atmospheres. These latter interactions, however, only take place with the upper atmosphere of the planet. The stellar wind or the stellar magnetic field can not directly initiate the outflow from the planet, but they interact with planetary atmospheres and help to confine them, or sometimes wipe them out, depending upon the stellar wind pressure \citep{Vidotto2020}. Mainly, it is the stellar radiation that heats up the atmosphere and drives a planetary wind against its gravity -- thus, studying the changes in the radiation from the host star is important to understand changes in atmospheric escape in close-in giant planets.

%importance of magnetic cycles
One particular temporal variation we explore in this paper is that of a stellar cycle.
The total radiation coming from the host star depends on its magnetic field. The foot point motions of magnetic flux ropes on the photosphere eventually lead to magnetic reconnection and transfer a huge amount of energy to the chromosphere and finally to the corona and beyond, which directly affects  close-in exoplanets \citep{Antolin2008,McIntosh2011}. If the host star is magnetically active, its radiation output is stronger during the peak activity period as is the case of our own Sun. A cyclic magnetic activity in the host star results in a cyclic variation in the radiation output of the star \citep{Yeo2014}, which can introduce a temporal variation in the physical properties of the planetary wind.

%our aim
In this paper, we investigate atmospheric escape in hot Jupiters orbiting stars with cyclic magnetic activities. For that, we implement a 1D hydrodynamic escape model that takes into account the effects of stellar photoionisation on planetary atmospheres. Given the difficulties in measuring stellar high-energy radiation (X-ray plus extreme ultraviolet, XUV from now on), we first consider a hypothetical system where a hot Jupiter orbits the Sun at a distance of 0.05 au. The direct measurements of XUV radiation for the Sun are available from space based data \citep{Woods2005}. We incorporate this observed cyclic variation of solar XUV flux in our model to investigate how this changes the physical properties of atmospheric escape in our fictitious planet. Although the X-ray part of the spectra has a small cross section for photo-absorption in our hydrogen-dominated  atmospheres, we use the full solar XUV  flux for all our solar calculations. However, direct measurements of the stellar XUV radiation are not available, because most of this energy is absorbed in the interstellar medium and upper atmosphere of the Earth. Some indirect ways to derive stellar XUV fluxes have been presented in the literature \citep{Ehrenreich2011,Youngblood2016, Peacock2019}, but here we use solar data to derive the XUV radiation from other stars. We do that by extrapolating the relation between the surface magnetic flux of the Sun and its XUV radiation over the available data of solar cycle 24. Using our newly derived formula, we infer the XUV radiation from the available magnetic field observations of HD189733 over the time span of a few years and investigate their effect on planetary escape rate and synthetic Ly$\alpha$ and H$\alpha$ transits.

%organization of the paper
The organization of the paper is as follows. In the next section, we discuss the relationship between magnetic fields and the high energy radiation of the Sun during its magnetic cycle.
Our hydrodynamic planetary escape model including a comparison study by considering a full spectral energy distribution and a monochromatic incident stellar radiation is presented in Section~\ref{sec:Non-mono}. In Section~\ref{sec:results}, we present our results on the effect of the solar magnetic cycle on the properties of a fictitious hot-Jupiter atmosphere. How to use the magnetic field of a star to probe its XUV radiation is given in Section~\ref{sec:xuv}. The calculated XUV radiation from the magnetic field of HD189733 from different epochs, its effect on the planetary atmosphere, including synthetic transit spectra, are discussed in Section~\ref{sec:HD189733}. Finally our conclusions are summarized in Section~\ref{sec:conclusion}.

\section{Relationship between magnetic cycle and  XUV radiation}\label{sec:cycle}
Magnetic field is a ubiquitous property of stars and many of them show a cyclic magnetic activity. The underlying dynamo mechanism i.e., the non-linear interaction between velocity field and magnetic field in the convection zone, drives the cyclic magnetic activity in  stars \citep{Parker55a,CSD95, Charbonneau14, HKC14, HCM17,Hazra19}. The CaII H\&K project of the Mount Wilson Observatory \citep{Wilson78,Noyes84a,Baliunas95} has monitored chromospheric activity of  111 stars of spectral types F2-M2 on or near the main sequence, and found that Sun-like cyclic activity is common and exhibited by many cool dwarfs. Recently \citet{BoroSaikia18} analysed chromospheric activity of 4454 cool stars from a combination of archival HARPS spectra and multiple others surveys, and concluded that cyclic activity is a common feature of others stars as well. Because of exoplanet detection biases,  most of the host stars of discovered exoplanets are magnetically less active, however, cyclic magnetic activities are detected in a few of them (e.g., $\epsilon$ Eridani, Tau Boo, HD189733). The evidence of a magnetic cycle in $\epsilon$ Eridani from CaII H \& K lines was reported in \citet{Gray1995} and recently confirmed by \citet{Metcalfe2013} with a period of 3 years. From spectropolarimetric data, \citet{2008MNRAS.385.1179D} discovered a magnetic cycle  in Tau Boo, a star that hosts Tau Boo b, a gas giant planet with orbital period of 3 days. Tau Boo shows a magnetic cycle with a period of 1 year or less \citep{Fares09, 2016MNRAS.459.4325M, 2018MNRAS.479.5266J}. A variable magnetic activity has also been detected in the planet-hosting star HD189733 but a regular cyclic activity is not yet confirmed \citep{Fares17}.

In view of the fact that cool dwarf stars are, overall, magnetically active and they might have a cyclic activity, here
we want to study how the cyclic magnetic activity of host stars would affect the dynamics of exoplanetary atmospheres. One of the main factors that affects atmospheric escape is the XUV radiation from the host star. Although it is generally accepted that the EUV (extreme ultra-violet) radiation is responsible for the heating of atmosphere by photoionization, there is a little consensus about atmosphere heating due to X-ray radiation. \citet{Cecchi-Pestellini06} study the X-ray heating of planetary atmospheres considering photoionisation of hydrogen and helium, and concludes that X-ray heating is important for young stars. \citet{Penz08} find that X-ray heating can evaporate most of the planets within the orbital distance of 0.05 au depending upon the strength of X-ray radiation. If we consider that the atmosphere of an exoplanet is dominated by hydrogen, then the X-ray part of heating hardly plays any role, because the hydrogen scattering cross section responsible for photoionisation is very small in the X-ray range in comparison to the EUV range \citep{Verner96,Bzowski13}. However, X-ray heating can be important if we consider  other elements in the atmosphere, and hence overall the whole XUV range is usually believed to play an important role in evaporating planetary atmospheres.

\begin{figure*}
\includegraphics[width=1.0\textwidth]{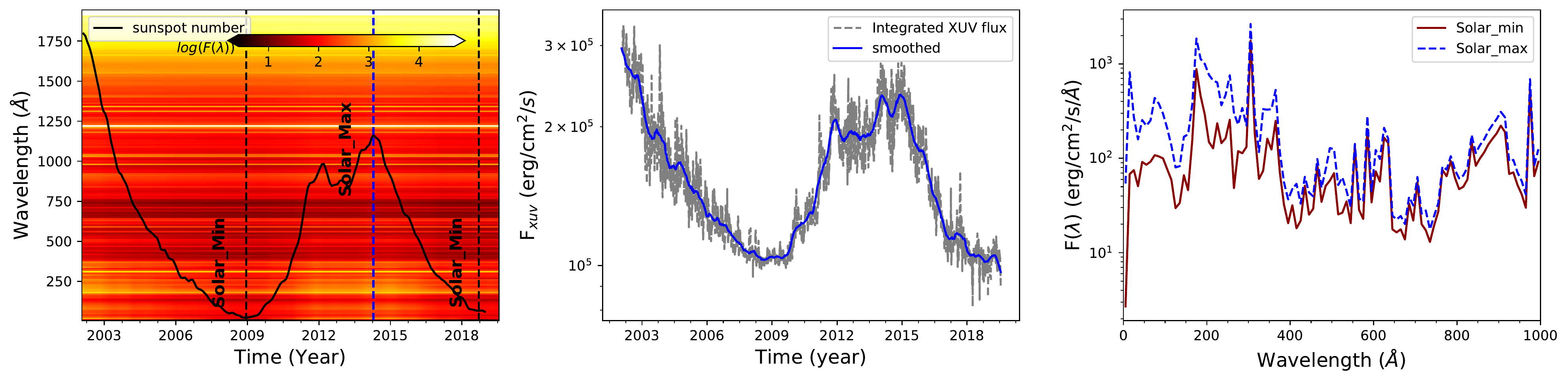}
\caption{Solar XUV spectra over a time span of a few years [2002 -2019]. (a) Six months averaged of full XUV spectra (5-1940 $\Angstrom$) is plotted with time, with colour denoting
the log of the spectral energy density (F$(\lambda)$), given in units of erg/cm$^2$/s/$\Angstrom$. The black solid line shows the sunspot cycle for the same time interval with sunspot number ranges from 3 to 180. The vertical black dashed lines mark the solar minima and blue vertical dashed line shows solar maximum.  (b) The variation of the XUV flux (F$_{\rm xuv}$) integrated over the wavelength range of 5-915 $\Angstrom$ as a function of time (grey dashed line) and its six-month smoothed curve (blue solid line). (c) Solar SED over the XUV wavelength range at two different times: blue dashed line shows the spectra during a solar maximum and red solid line represents the spectra during a solar minimum.}
\label{fig:spectra}
\end{figure*}

The stellar XUV radiation is highly modulated by the stellar magnetic field. Since measuring magnetic field and XUV radiation simultaneously from a distant planet-hosting star is very challenging, we first focus on the Sun as a case study.
As the EUV part of the solar spectra gets absorbed by the upper atmosphere of the Earth, we can not measure it from ground-based facilities and we did not have a long-term data available for EUV spectra, until recently with the advancement of space-based observatories, making it now possible to measure the EUV radiation of the Sun. Here, we have taken data from the Solar Extreme Ultraviolet Experiment (SEE) instrument on the NASA Thermosphere Ionosphere Mesosphere Energetics Dynamics (TIMED) spacecraft, which is available from January 2002 to present  \citep{Woods2005}. Since we are interested in the long term effect of cyclic magnetic activity on atmospheric escape, we use the flare-removed data from the SEE instrument. The whole available half-yearly averaged XUV spectra is shown in Fig. ~\ref{fig:spectra}(a). We over plot the yearly averaged sunspot numbers to see the correlation of the spectra with the solar cycle. Two solar minima are shown using vertical black dashed lines and solar maximum for the cycle 24 is shown using blue dashed line.

We integrate the whole spectra over the XUV wavelength range 5--915~\AA\ to get the XUV flux, $F_{\rm xuv} = \int_5^{915 \textrm{\AA}} F(\lambda)d\lambda$, where $F(\lambda)$ is the Spectral Energy Distribution (SED) over the XUV wavelength range. The cyclic modulation of the whole XUV spectra becomes more clear when we plot the integrated F$_{\rm xuv}$ as a function of time, as shown in Fig.~\ref{fig:spectra}(b). The grey dashed line shows the daily $F_{\rm {xuv}}$ and the blue solid line shows the six-month smoothed curve for F$_{\rm xuv}$, which follows the similar sunspot number curve. The whole XUV spectra during the solar maximum and solar minimum are plotted in Fig.~\ref{fig:spectra}(c). A clear distinction between different spectral intensity during solar maximum and solar minimum is evident from this figure over the whole XUV wavelengths with noticeable differences  found for high energy spectra below 600 \AA\ (Fig.~\ref{fig:spectra}(c)). Therefore, it becomes clear that the solar XUV spectra follows the nature of the solar cycle and it becomes strong during solar maximum and weak at solar minimum. We utilize this time distribution of XUV spectra to study how the cyclic magnetic activity of the host star affects the characteristics of its planetary atmosphere.

\section{Hydrodynamic escape model}\label{sec:Non-mono}

Our hydrodynamic escape model is based on the 1D planetary wind model that has recently been presented by \citet{Allan2019} which is similar to the model presented in \citet{Murray-clay2009}. These authors, however, assume stellar photons have energy concentrated in one single wavelength, while in our model, we assume the energy is distributed according to the solar SED. One important criterion assumed in these models is the fluid condition of planetary atmospheres. As observation suggests dense escaping atmospheres in close-in planets \citep{Ehrenreich2011}, planetary atmospheres are likely to be collisional and the fluid approximation remains valid. The Knudsen number, $K_n = \lambda_{\rm mfp}/h$, where $\lambda_{\rm mfp}$ is the mean free path of the atmospheric particle and $h$ is the atmospheric scale height, is a very good indicator to validate whether the atmosphere is collisional or not. Planetary atmospheres can be depicted as collisional in case $K_n$ $<<$ $1$ and the equation of fluid dynamics can be applied to model planetary atmospheres. In our model, we solve the equations of fluid dynamics for the escaping planetary atmosphere in a co-rotating frame including ionisation balance. After the simulation converges, we calculate the Knudsen number as well to make sure the fluid approximation is valid throughout our simulation domain.

In steady state, the momentum and energy equations are given below:
\begin{equation}
u\frac{du}{dr} = -\frac{1}{\rho}\frac{dP}{dr} - \frac{GM_{\rm pl}}{r^2} + \frac{3GM_{\star}r}{a^3} \label{eq:mom}
\end{equation}
\begin{equation}
\rho u \frac{d}{dr}\left[\frac{k_B T}{(\gamma -1)m}\right] = \frac{k_BT}{m}u\frac{d\rho}{dr} + Q - C, \label{eq:energy}
\end{equation}
where $u$, $\rho$, $P$ and $T$ are the planetary wind velocity, mass density, thermal pressure and temperature respectively. $k_B$ is the Boltzman constant. $r$ is the radial distance from the centre of the planet and $a$ is the star-planet distance. $M_{\star}$ and $M_{\rm pl}$ are the mass of the host star and planet respectively.
We take $\gamma$ = 5/3, which is the ratio of specific heats for a monatomic ideal gas, and $m$ is the mean particle mass. The first term on the right hand side of  equation~(\ref{eq:mom}) is the pressure gradient term and the second term incorporates the gravitational attraction of the planet. The third term (i.e., the tidal term) represents the resultant effect of centrifugal force and differential stellar gravity along the ray between the planet and the star. In the energy equation (equation~(\ref{eq:energy})) the term on the left, i.e., the change of internal energy of the atmosphere, is being balanced by the successive heating and cooling terms on the right. The first term on right represents the cooling due to expansion of the atmospheric gas, Q is the heating due to incident stellar radiation and C represents the cooling term due to Ly$\alpha$ cooling. Ly$\alpha$ cooling occurs due to the loss of radiation emitted by the neutral hydrogen which is collisionaly excited by electrons. The volumetric Ly$\alpha$ cooling rate is $C = 7.5 \times 10^{-19} n_+n_0exp[-1.183\times 10^5/T]$, where $n_+$ is the number density of protons equivalent to number density of electron.

In case the incident stellar radiation is distributed over a SED $F(\lambda)$, the heating term can be written as
\begin{equation}\label{eq:heat}
Q = \int_{\lambda_1}^{\lambda_2}\frac{F(\lambda)}{\lambda_0}(\lambda_0 - \lambda)n_0\sigma(\lambda)e^{-\tau(\lambda)}d\lambda
\end{equation}
where $\lambda_1$ is the extreme high energy wavelength limit of the XUV spectra and $\lambda_2$ is the low energy part of the spectrum. $\lambda_0$ = 912.0 $\Angstrom$ is the threshold wavelength to ionize hydrogen atom. In all of our calculation, the range of XUV radiation is taken as 5-915 $\Angstrom$ and $\lambda_1$ and $\lambda_2$ are set at 5 $\Angstrom$ and 915 $\Angstrom$, respectively. Note that we set the upper limit $\lambda_2$ = 915 $\Angstrom$ instead of 912 $\Angstrom$, as our incident stellar radiation is equally distributed with an interval of 5 $\Angstrom$. $n_0$ is the number density of neutral hydrogen, $\sigma(\lambda)$ is the wavelength-dependent scattering cross section of hydrogen atoms, and $\tau(\lambda)$ is optical depth that varies with wavelengths. The expression for $\sigma(\lambda)$ is taken from \citet{Bourrier2016}, which is based on \citet{Verner96, Bzowski13} as given below
\begin{equation}
  \sigma(\lambda) = 6.538 \times 10^{-32} \left(\frac{29.62}{\sqrt{\lambda}} +1\right)^{-2.963}(\lambda - 28846.9)^2\lambda^{2.0185}.
\end{equation}
The unit of $\sigma(\lambda)$ is cm$^2$ and the wavelength is given in $\Angstrom$. The wavelength-dependent optical depth along the path of stellar radiation can be written as
\begin{equation}\label{eq:tau}
 \tau(\lambda) = \int_{\lambda_1}^{\lambda_2}\int_{\infty}^r n_0\sigma(\lambda)drd\lambda.
\end{equation}
 After the stellar XUV radiation gets absorbed in the planetary atmosphere, it ionises the neutral hydrogen depending upon its energy, and the rate of photoionisation must be balanced by the rate of radiative recombination and the advection rate of ions in order to reach an ionization equilibrium. The rate equation considering all these three ionisation processes is
\begin{equation}\label{eq:rate}
n_0\int_{\lambda_1}^{\lambda_2} \frac{1}{hc}F(\lambda)e^{-\tau(\lambda)} \sigma(\lambda) \lambda d\lambda  =  n_+^2\alpha_{rec} + \frac{1}{r^2}\frac{\partial}{\partial r}(r^2n_+u),
\end{equation}
where $n_+$ is the number density of ionized hydrogen and $n = n_+ + n_0$ is the total number density of hydrogen nuclei. $\alpha_{\rm rec} = 2.7 \times 10^{-13} (T/10^4 K)^{-0.9}$ is the recombination coefficient for protons and neutrals.
If we define the ionization fraction $f_+$ = $n_+/n$, the advective term can be rewritten, with the aid of the mass conservation equation ($n u r^2 = $constant), as
\begin{equation}\label{eq:advect}
    \frac{1}{r^2}\frac{\partial}{\partial r}(r^2n_+u) = nu \frac{\partial f_+}{\partial r} .
\end{equation}
Plugging equation~(\ref{eq:advect}) into the equation~(\ref{eq:rate}), the final rate equation in terms of ionization fraction can be put as following:
\begin{equation}\label{eq:final_rate}
\frac{\partial f_+}{\partial r} = \frac{1-f_+}{u}\int_{\lambda_1}^{\lambda_2} \frac{1}{hc}F(\lambda)e^{-\tau(\lambda)} \sigma(\lambda) \lambda d\lambda - nf_+^2\alpha_{rec}/u.
\end{equation}
Also, from mass continuity we have,
\begin{equation}\label{eq:continuity}
  \frac{\partial}{\partial r}(r^2\rho u) = 0.
\end{equation}

In order to find the structure of the planetary outflow, we need to solve the momentum equation (equation~(\ref{eq:mom})), energy equation (equation~(\ref{eq:energy})) and rate equation (equation~(\ref{eq:final_rate})) simultaneously following the mass conservation (equation~(\ref{eq:continuity})). We can see that the heat and rate equations (equations~(\ref{eq:heat}) and (\ref{eq:final_rate})) are written here considering the incident stellar radiation is distributed over a wide range of wavelengths. We call this  the  `non-monochromatic' case, to differentiate it from the `monochromatic' case, in which we assume the total integrated XUV flux, F$_{\rm xuv}$, is concentrated on a single wavelength ($\lambda_{\rm mono}$). In the monochromatic case, the heat equation is reduced to $Q_{\rm mono} = \epsilon F_{\rm xuv} e^{-\tau_{\rm mono}}\sigma_{\rm mono}n_0$. Unlike the non-monochromatic case, the optical depth $\tau_{\rm mono}$ = $ \sigma_{\rm mono}\int_{\infty}^r n_0 dr$ is now independent of wavelength, where $\sigma_{\rm mono}$ is the area of cross section for the monochromatic wavelength and $\epsilon$ takes care of the excess energy that contributes to the ionisation of photoelectrons and hence the acceleration that leads to the heating of the atmosphere. In case the incident radiation has a frequency $\nu_{\rm mono} = c/\lambda_{\rm mono}$, where $c$ is the speed of light, the measure of excess energy can be written as $\epsilon = (h\nu_{\rm mono} - 13.6 {\rm eV})/h\nu_{\rm mono}$. Several models of planetary wind \citep{Murray-clay2009, Allan2019} assume the XUV flux is concentrated at one photon energy $h\nu_{\rm mono}$ = 20 eV which gives $\epsilon = 0.32$. Similarly the rate equation (equation~(\ref{eq:rate})) converges to the equation~(3) of \citet{Allan2019} for the monochromatic case.

In our model, the planetary outflow originates from the substellar point on the planet and then subsequent mass loss of atmosphere occurs in the form of steady, hydrodynamic transonic wind. The escape rate is calculated by rendering our wind solution over all 4$\pi$ steradians as $\dot{M}$ = 4$\pi r^2 \rho u$. This escape rate must be taken as a upper limit on the total rate of photoevaporative mass loss because our model does not consider variation in the illumination of the day to night side of the planet. The initial subsonic flow gets accelerated transonicaly until it reaches an asymptotic steady speed which is known as terminal velocity ($u_{\rm term}$). The convergence of our simulation is estimated by calculating escape rate and terminal velocity of two subsequent runs that come under a certain limit. We assume the convergence is reached in our model when escape rate and terminal velocity for two subsequent runs are below 1\%.

\begin{figure}
\includegraphics[width=\columnwidth]{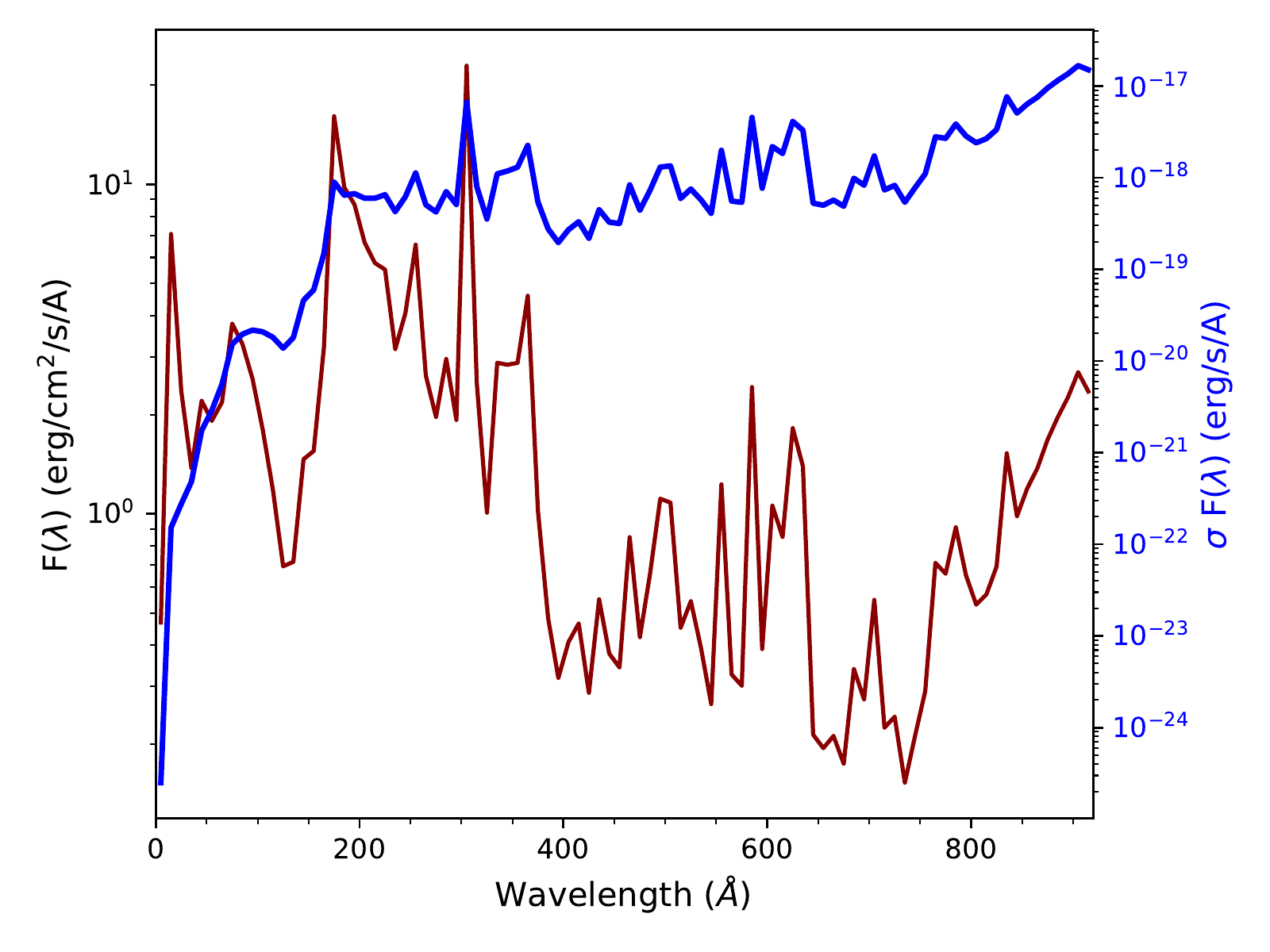}
\caption{XUV spectral energy distribution is shown in solid red line during solar maximum, April 2014 at 0.05 au. The blue solid line shows the product of the solar XUV flux and wavelength-dependent scattering cross section of hydrogen. }
\label{fig:sed}
\end{figure}

\subsection{Non-monochromatic vs monochromatic incident stellar radiations}
Here we compare whether a non-monochromatic spectrum has a distinct effect on atmospheric escape than the case of a monochromatic incident stellar radiation. We assume a hypothetical exoplanetary system consisting of a  hot Jupiter orbiting a Sun-like star with mass 1.0 M$_\odot$, and similar solar XUV spectrum. The orbital distance, mass and radius of the hot Jupiter are taken  as $0.05$ au, $0.7$ $M_{\rm Jup}$ and 1.4 $R_{\rm Jup}$ respectively.

\begin{figure}
\includegraphics[width=\columnwidth]{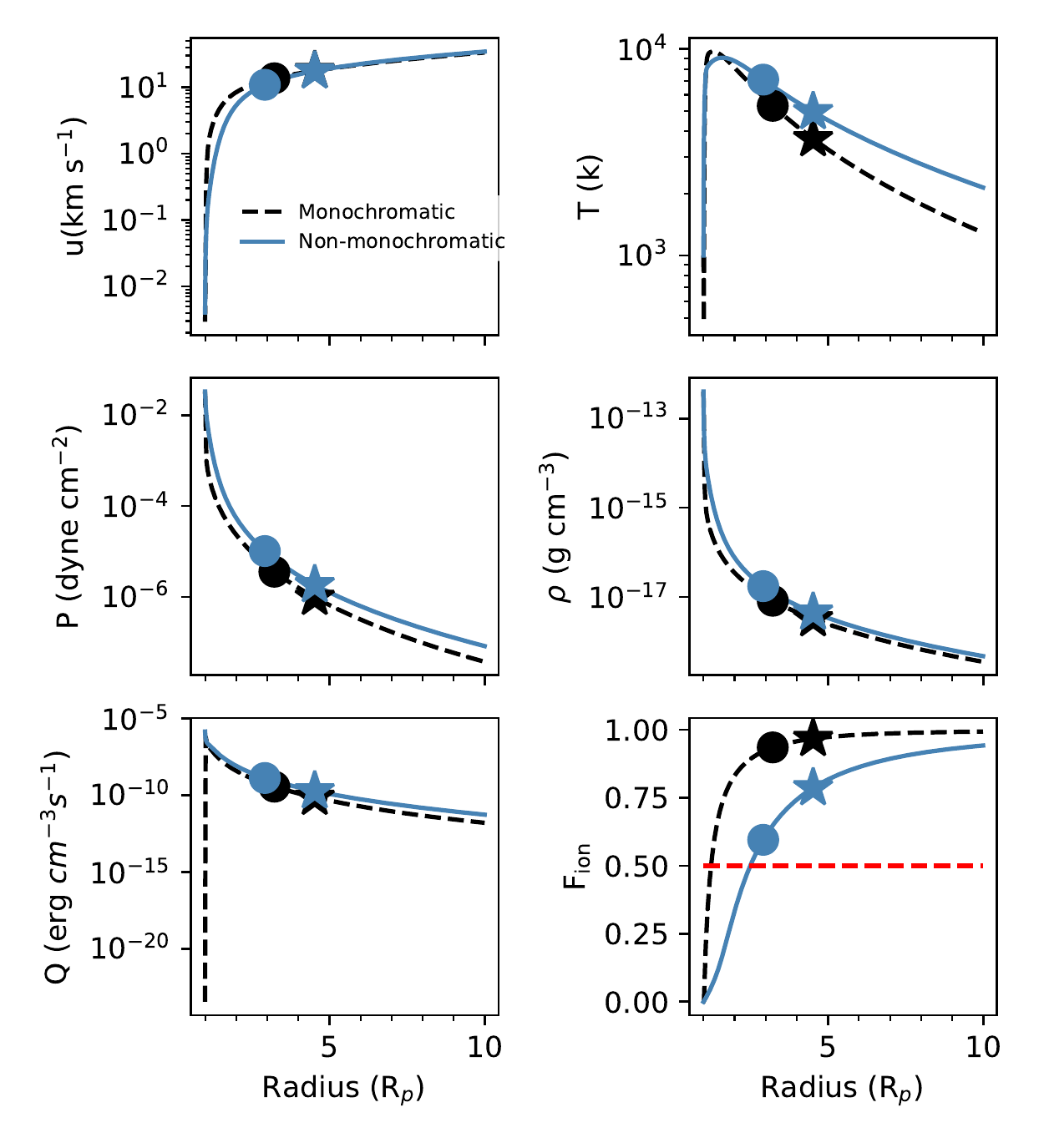}
\caption{Standard planetary wind (atmospheric escape) solution of a hot Jupiter system, assuming the Sun as its host star. Radial variations of basic atmospheric properties are shown. In the top panel, velocity (u) and temperature (T) of the wind are shown in the left and right. Similarly in the middle row, pressure (P) and mass density ($\rho$) are presented. Total volumetric heating rate and fraction of ionized hydrogen are shown in the bottom row in the left and right, respectively. The red horizontal dashed line in last panel marks  50\% of ionisation fraction. Solid blue and dashed black lines represent the non-monochromatic and monochromatic stellar XUV radiation. Radial distances are always plotted in units of planetary radii R$_p$.}
\label{fig:wind_sol}
\end{figure}

The red solid line in Fig.~\ref{fig:sed} shows the SED from the Sun, calculated at the orbital distance of the planet,  in the XUV wavelength range (5-915 $\Angstrom$) during solar maximum at April 2014, as explained in Section~\ref{sec:cycle}. This  SED spectrum is incorporated in our escape model. The solutions with the non-monochromatic F$(\lambda)$ are shown in Fig.~\ref{fig:wind_sol} using blue solid lines. These results are qualitatively  similar to the results presented earlier in the literature \citep{Murray-clay2009,Allan2019}. The stellar XUV flux drives a transonic wind with a maximum temperature of 10$^4$ K and velocity almost $\sim$ 10 km s$^{-1}$ as shown in the top row of Fig.~\ref{fig:wind_sol}. Pressure and density profiles of the wind with radius are shown in the middle row of the same figure. The heating profile and corresponding fraction of ionised hydrogen are presented in the bottom row. The Roche lobe boundary, R$_{\rm roche}$ = $a\left(\frac{M_p}{3M_\star}\right)^{1/3}$, where stellar gravity balances the planetary gravity is shown using the star marker in each plot. Sonic points where wind velocity is same as the sound speed are shown using the filled circle. Note that the sonic points are within the Roche lobe boundary.

To compare with the non-monochromatic case, we perform another simulation
assuming that the whole integrated SED is concentrated in the photon energy h$\nu_{\rm mono}$ = 20 eV. The aim of performing this simulation is to understand whether assuming the whole SED concentrated in a particular monochromatic wavelength is critically different from considering the full SED.
To be precise, in the monochromatic case, we take F$_{xuv}$ = $\int_{5}^{915 \Angstrom} F(\lambda)d\lambda$ = 1968.94 erg cm$^{-2}$ s$^{-1}$ at the planet, where F$(\lambda)$ is the SED as shown in Fig.~\ref{fig:sed} by the solid red line. This monochromatic assumption is not too  bad considering that the scattering cross section $\sigma(\lambda)$ is very small at the high-energy part of the radiation. In this case, most of the high-energy photon enters deep into the planetary atmosphere without ionizing the hydrogen atom. This scenario is evident from the blue solid line plot of ($\sigma \times F(\lambda)$) vs $\lambda$ in Fig.~\ref{fig:sed}. We see that $\sigma \times F(\lambda)$ sharply falls for the wavelengths below 200 $\Angstrom$.

The results of this simulation (monochromatic case) are shown by the dashed black lines in Fig.~\ref{fig:wind_sol}. These results are not too  different from the results we obtain by incorporating the non-monochromatic radiation distribution (solid blue lines). The most notable differences are found in the temperature and ionization fraction plots. For the non-monochromatic case, the temperature of the wind remains slightly higher but the ionization fraction of hydrogen remains smaller than the monochromatic case. The reason for this is that in the non-monochromatic case, photons of different spectral energies penetrate into different depths of the atmosphere, giving rise to an extended heating profile in altitude and hence a higher temperature at high altitudes. An abrupt initial increase of ionisation fraction in the monochromatic case is due to the steep temperature increase near the base of the atmosphere, compared to the more extended temperature profile of the non-monochromatic case (see right panel of top row in Fig.~\ref{fig:wind_sol}). The ionisation rate of hydrogen  depends on how the photoionisation rate is being balanced by the radiative recombination rate as well as the rate at which ions are advected upwards into the atmosphere. A peak in temperature very close to planet in the monochromatic case ionises more hydrogen near the base of the atmosphere compared to the non-monochromatic case, but other processes, i.e., advection of ionised hydrogen upwards in the atmosphere and the negligible recombination rates are comparable for both cases.

Similar results have  also been reported in \citet{Koskinen13} who found an overestimation of the ionisation fraction for the monochromatic case. They also found that ionization transition of hydrogen atom happens at higher altitude for the non-monochromatic case than for the monochromatic case. The horizontal dashed red line in the bottom right panel of Fig.~\ref{fig:wind_sol} shows where the ionisation fraction reaches 50$\%$ in our simulation. We see similar result as \citet{Koskinen13} that for non-monochromatic case, this ionisation transition occurs at a  higher atmospheric height than for the monochromatic case.

\begin{figure}
\includegraphics[width=\columnwidth]{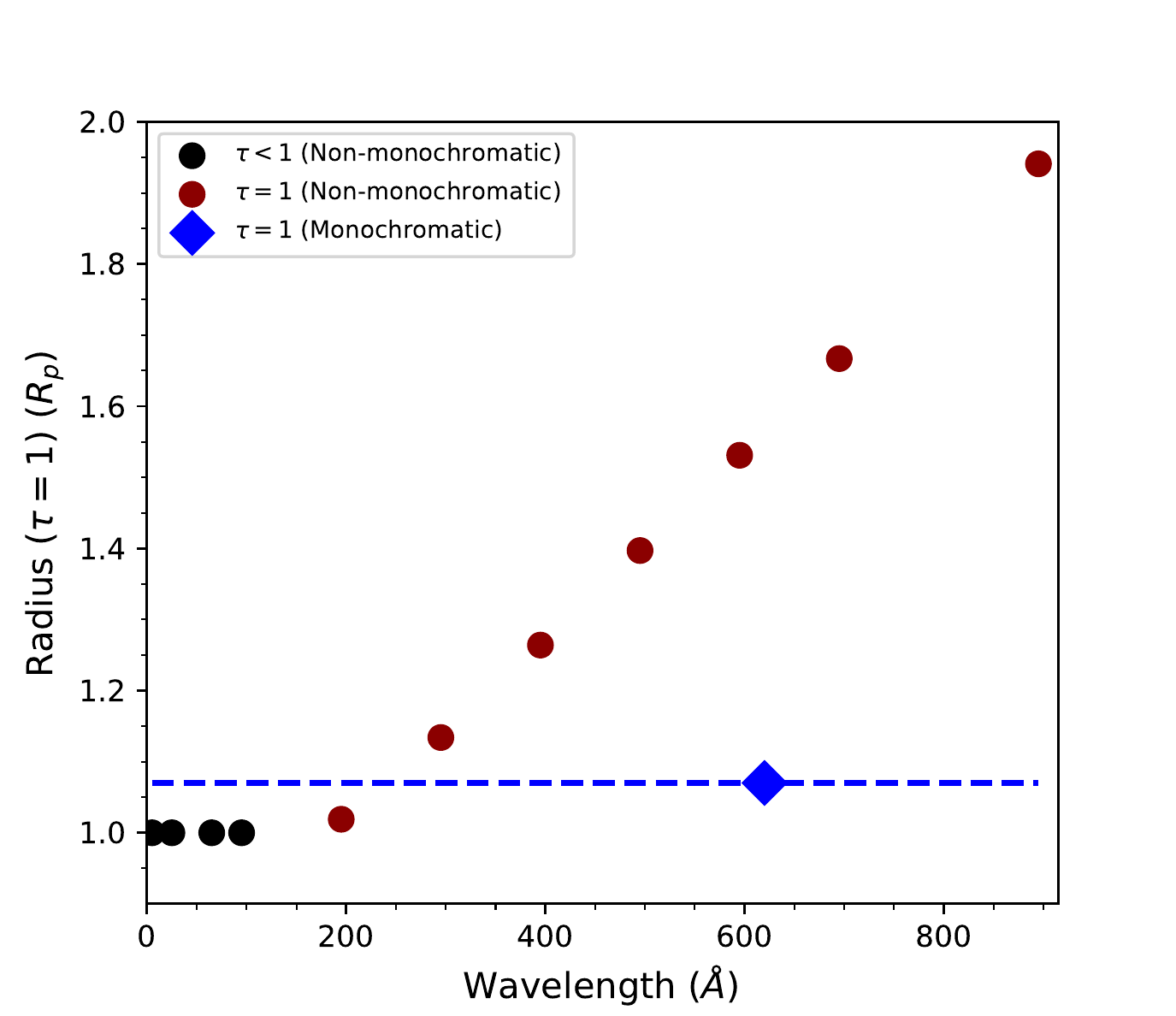}
\caption{Radius in the planetary atmosphere, where the optical depth becomes one, are shown for non-monochromatic and monochromatic cases. The blue dashed line shows the monochromatic $\tau$ = 1 radius which is always a constant surface but for non-monochromatic case, it varies a lot as shown by red filled circles. The black filled circles show the wavelength for which planetary atmosphere is always optically thin.  Radius are given in the units of R$_p$.}
\label{fig:tau}
\end{figure}

In Fig.~\ref{fig:tau}, we plot the radius of planetary atmosphere where the optical depth, $\tau$ is 1 as function of wavelengths for both the non-monochromatic and monochromatic cases. For the non-monochromatic case, the atmosphere becomes increasingly opaque as we go to  higher wavelengths. The radius of the planetary atmosphere where $\tau = 1$ (i.e., a fraction $e^{-\tau}$ = 1/e = 37$\%$ of the incident photons is absorbed) is higher for the longer incident wavelength as it is clear from the dark red filled points in the plot. The optical depth for the very short wavelength of the spectra (high energy photons) does not even reach one and the atmosphere remains optically thin for those part of the spectrum (black filled circles in the Fig.~\ref{fig:tau}). As we mentioned earlier, the different cross sections of hydrogen  over different wavelengths (see the blue curve on Fig.~\ref{fig:sed}) allow photons to penetrate at different depths in the atmosphere resulting in different optical depths. On the contrary, in the monochromatic case, all of the XUV photons get absorbed near the base of the atmosphere and $\tau = 1$ is reached at r = 1.07 R$_p$. If we could observe the planetary atmosphere in 912 $\Angstrom$ in which the absorption cross section peaks, it would look puffy and radius of planetary atmosphere where $\tau=1$ would be almost twice the size of the planet geometric disc.

We also find that the of escape rates for two cases are quite similar: 2.1 $\times$ 10$^{11}$ g s$^{-1}$ and 1.5 $\times$ 10$^{11}$ g s$^{-1}$ for the non-monochromatic case and monochromatic case, respectively. Hence, although there are a few differences (e.g., different temperature profiles, f$_{\rm ion}$ and different radius of the $\tau = 1$ surface) in the results obtained by considering two cases, qualitatively both of them behave in similar ways. The key differences between the non-monochromatic  and monochromatic models are summarised in Table~\ref{tab:nonmono}.

\begin{table*}
	\centering
	\caption{Key differences in modelling the incident stellar radiation as non-monochromatic (SED) and monochromatic on calculations of the escaping atmosphere of a hot Jupiter.}
	\label{tab:nonmono}
	\begin{tabular}{|l|c|c|} % four columns, alignment for each
		\hline
		Parameters & Non-monochromatic case & Monochromatic case\\
		\hline
    Incident radiation &  $F(\lambda)$ distributed over $\lambda_1$ = 5 $\Angstrom$ to $\lambda_2$ = 915 $\Angstrom$ & $F_{\rm xuv} = \int_{\lambda_1}^{\lambda_2} F(\lambda)d\lambda$ is concentrated on $\lambda= \lambda_{\rm mono}$\\
		Absorption cross section & $\sigma(\lambda) = 6.538 \times 10^{-32} \left(\frac{29.62}{\sqrt{\lambda}} +1\right)^{-2.963}(\lambda - 28846.9)^2\lambda^{2.0185}$ cm$^2$ & $\sigma_{\rm mono}$ = 1.89 $\times$ 10$^{-18}$ cm$^2$\\
    & & \\
    Optical depth & $\tau(\lambda)$ = $\int_{\lambda_1}^{\lambda_2}\int_{\infty}^rn_0\sigma(\lambda)drd\lambda$ & $\tau_{\rm mono}$ = $\sigma_{\rm mono}\int_{\infty}^r n_0 dr $ \\\\

    Photoinization heating & $Q = \int_{\lambda_1}^{\lambda_2}%\frac{(\lambda_0 - \lambda)}{\lambda_0}
    \epsilon(\lambda){F(\lambda)}e^{-\tau(\lambda)}\sigma(\lambda)n_0 d\lambda $  &  $ Q_{\rm mono} = \epsilon F_{\rm xuv} e^{-\tau_{\rm mono}} \sigma_{\rm mono} n_0$ \\
    & &  \\
    %Efficiency
    Excess photon energy & $\epsilon(\lambda) = \frac{h\nu-h\nu_0}{h\nu} =  \frac{\lambda_0-\lambda}{\lambda_0}$, $h\nu_0 = \frac{hc}{\lambda_0} = 13.6 eV$   & $\epsilon = 0.32$, with $\lambda = \lambda_{\rm mono}$ so that $hc/\lambda_{\rm mono} = 20eV$\\ \\
    $\tau$ = 1 surface & Varies with wavelengths & Remains constant\\ \\
    Ionization at 50\% level & Transition occurs at  high altitude &  Transition occurs close to the planet \\ \\
    Temperature  & An extended profile &  A steep profile close to planet then decays \\
		\hline
	\end{tabular}
\end{table*}

To illustrate the atmospheric variability of our fictitious hot Jupiter over the solar cycle, we  calculate the atmospheric properties at two times during solar maximum and solar minimum, considering both the non-monochromatic and monochromatic cases. The results for both cases are not significantly different in terms of variability of atmospheric properties at the maximum and minimum epochs, and the non-monochromatic case takes more computation time as expected. Therefore, for the rest of our study, we use the monochromatic approximation of incident stellar XUV radiation concentrated on photon energy h$\nu_{\rm mono}$ = 20 eV.

\section{Variability of atmospheric escape in hot Jupiters over stellar cycle}\label{sec:results}
In this section, we investigate  variability of atmospheric escape in our fictitious planet, by considering  the  evolution of stellar radiation over a magnetic cycle. As we mentioned earlier, our hypothetical exoplanetary system consists of a hot Jupiter of mass 0.7M$_{\rm Jup}$ orbiting the Sun as its host star. This planetary system is similar to the HD209458 exoplanetary system. As the cycle evolves, we implement the corresponding XUV radiation integrated over 5-915 $\Angstrom$ wavelength range as an input in our 1D planetary model (Section~\ref{sec:Non-mono}). Note that here we are using a monochromatic approximation of incident XUV flux, i.e., the entire XUV spectra is concentrated on a photon energy of 20 eV. At each of the simulated epochs during the solar cycle, we take F$_{\rm xuv}$ = $\int_{5}^{915 \Angstrom} F{(\lambda)}d\lambda$, where F$(\lambda)$ is the SED as shown in Fig.~\ref{fig:spectra}(a).

\begin{figure*}
\includegraphics[width=0.95\textwidth]{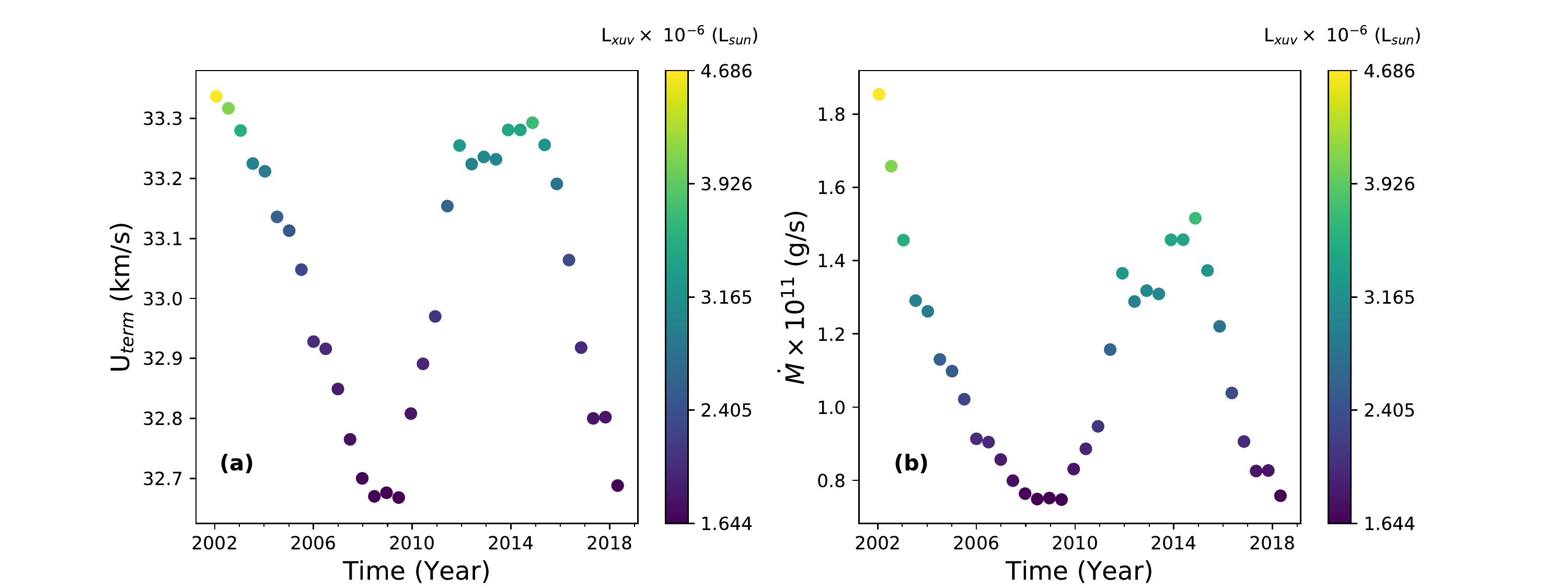}
\caption{(a) Terminal velocity of the escaping atmosphere for different incident solar XUV fluxes spanning one and a half solar cycle. (b) Same as (a) but for escape rate. The luminosity of the incident solar XUV radiation is shown in the colorbar.}
\label{fig:mdot}
\end{figure*}

\subsection{Cyclic variation of the outflow properties}\label{sec:cycle_outflow}
We consider 34 cases where \Fxuv\ is chosen at different epochs spanning one and a half magnetic cycles with an interval of 6 months starting from January 2002. By keeping all other parameters the same, we run our atmospheric escape model with these different \Fxuv\ values and investigate how the planetary wind properties vary along the stellar magnetic cycle. Photoionisation due to the incident F$_{\rm xuv}$ is mainly responsible for driving the transonic wind and when F$_{\rm xuv}$ changes with the magnetic cycle, the corresponding photoionisation changes and so the properties of the outflow. We find that the radial profiles of temperature, pressure, density and velocity are higher for stronger incident XUV radiation and lower for the weak radiation. The  evolution of terminal velocity and escape rate are shown in Figs.~\ref{fig:mdot}(a) and (b) respectively. The corresponding XUV luminosity is also shown using the colorbar. During cycle maximum, strong XUV radiation leads to higher velocity and higher density of the wind than at cycle minimum phase and allows the planetary wind to remove more material from the planetary atmosphere (Fig.~\ref{fig:mdot}(b)). A factor of 2.9 increment in the incident XUV flux between solar maximum and solar minimum increases the escape rate by a factor of 2.5. We see a similar cyclic variation in the terminal velocity over the magnetic cycle. However, the variation in the terminal velocity is not as significant as the escape rate. The terminal velocity is defined as the velocity of atmospheric outflow that reaches an asymptotic value and in our simulations this velocity is achieved around 10R$_p$.
The stellar radiation gets absorbed very deep in the atmosphere, and drives a bulk outflow that become super-sonic and eventually escapes the planet. Hence a strong F$_{\rm xuv}$ penetrates deep into the atmosphere initiating a bulk outflow very close to the planetary surface with a high temperature. This results in a high sonic speed. For a weak \Fxuv , the outflow attains sonic speed at a higher altitude with a low value. In both cases, the external heating penetrates deep in the atmosphere, affecting the thermal pressure gradient force in the sub-sonic region. Forces or heating applied in the sub-sonic region affect more significantly the mass-loss rate than the terminal velocity (For having a larger change in terminal velocity, the force/energy deposition should be applied in the super-sonic part of the outflow).
As a result, the cyclic variation of \Fxuv\ does not contribute a significant variation in the terminal velocity.

Direct measurement of XUV flux for the Sun is only available for a few years starting from 2002 covering the decaying part of cycle 23 and full cycle 24. The cyclic variation of XUV radiation is not strong in these cycles resulting in a relatively small variation in the rate of the planetary escaping material.  If we observe  the behaviour of the solar cycle from the last 400 years of available record of sunspot numbers \citep{SIDC14}, we see that the solar cycle varies a lot in their strength, with cycle 19 being the strongest one, while cycle 24 is one of the weakest. Reconstructed solar EUV radiation data from geomagnetic record \citep{Svalgaard2015} provides an opportunity for direct comparison of EUV radiation with solar cycle, which suggests that EUV radiation is highly correlated with the solar cycle (see Fig. 22 of \citet{Svalgaard2015}). The reconstructed EUV flux also suggests that the EUV flux reaches the same low value at every sunspot minimum possibly including grand minima as well. Hence the stronger solar cycle would have more EUV radiation compared to the weakest one. A factor of 2.9 between XUV fluxes during maximum and minimum in cycle 23-24, could have been much larger if we would have considered another cycle (e.g., cycle 19) and as a consequence a stronger cyclic variation of escape rate of the planetary atmosphere would be expected.

\subsection{Effect of cyclic variation in spectroscopic transits}\label{sec:transit}
The influence of cyclic magnetic activity and corresponding cyclic behavior of XUV radiation on planetary escape will become visible if we calculate the spectroscopic transit in the hydrogen lines, the most abundant element in the planetary atmosphere. In particular, the most prominent stellar  line in the far ultraviolet, Lyman $\alpha$, helps to detect atmospheric escape from most of the gas-giants planets.

To confirm that the atmosphere is escaping from the planet, one needs to show that the atmosphere extends beyond the Roche lobe radius, overcoming the planetary gravity. This means that the outflow could eventually succumb to stellar gravity or being pushed out of the system by, e.g., radiation pressure forces or the interaction with the stellar wind. \citet{Vidal-Madjar2003} detected a large absorption of 15 $\pm$ 4\% in the stellar Ly$\alpha$ line of
HD209458 and concluded that this large absorption can only occur if the atmosphere extends beyond the Roche lobe of the planet, which provides a strong observational support of escaping atmosphere from the planet. On the other hand, H$\alpha$ absorption has also been found
in the transmission spectra of many exoplanets \citep{Winn04, Jensen12, Christie13, Barnes16}.
Although detection of Ly$\alpha$ has its own merit, detecting H$\alpha$ line would provide additional constraints on the density and temperature of the exoplanetary atmosphere as H$\alpha$ absorption has a strong dependence on temperature due to the collisional excitation rate of the n=1 to n=2 transition \citep{Christie13}.

The solutions of our hydrodynamic calculation of the planetary wind for the 34 cases mentioned earlier are used to calculate the spectroscopic transit in the Ly$\alpha$ and H$\alpha$ lines. Note that all 34 cases include different XUV fluxes over one and a half magnetic cycles of the Sun. To model the Ly$\alpha$ and H$\alpha$ lines we need to know the density of neutral hydrogen in ground state (n = 1) and first excited state (n = 2) respectively.
For that, we solve the statistical equilibrium equation in the coronal model approximation \citep{DelZanna18}, which takes temperature and electron density of the atmosphere as input parameters. These are obtained from our planetary wind calculations to derive the population density. In this approximation, the spontaneous radiative decay balances the excitation process but neglects the collisional de-excitation. The direct excitation from the ground state only is included and we use the Chianti software \citep{chianti9.0.1} in IDL to perform the direct excitation from the ground state to the first excited state.

We use a ray tracing method to simulate planetary transit as observed in both Ly$\alpha$ and H$\alpha$ as described in \citet{ Vidotto2018b, Allan2019}. We create a three dimensional Cartesian grid with 201 evenly spaced grid in each direction keeping the planet on its center. Each cell of the grid is filled with the velocity, density and temperature obtained from the one-dimensional hydrodynamic calculations of the planetary wind taking into consideration that Doppler shifts result from the projection of the wind velocities along the line of sight. One of our axis in the grid is aligned along the observer-star line and the other two axes appear as the plane-of-the-sky with 201 $\times$ 201 cells. We have used 51 velocity channels starting from u$_{\rm channel}$ = -500 km/s to 500 km s$^{-1}$ with an increment of 50 km/s in the outer velocities and an improved resolution at line center (35 channels are distributed over $\pm$ 100 km s$^{-1}$).

The frequency-dependent or velocity-dependent optical depth along a single ray in the direction connecting star-planet system to the observer can be written as
\begin{equation}\label{eq:tau_nu}
  \tau_\nu = \int n_0 \sigma_t \phi_\nu dz
\end{equation}
where $n_0$ is density of neutral hydrogen, $\sigma_t$ is the absorption cross-section at the line center and $\phi_\nu$ is the Voigt line profile. $\sigma_t$ is calculated as $\sigma_t = \pi e^2 f/m_ec$ where $f$ is oscillator strength, $e$ is the electron charge and $m_e$ is the mass of the electron. The oscillator strengths are different for different lines. We take $f = 0.416410$ for Ly$\alpha$ and $f = 0.64108$ for H$\alpha$. These values are taken from the NIST catalogue \footnote{\url{https://www.nist.gov/pml/atomic-spectra-database}} of atomic spectra data base. The Voigt line profile is taken as a standard Voigt profile (see equation~(12) of \citet{Allan2019}) by using IDL's inbuilt {\it voigt} function. We remind the reader that the optical depth, $\tau_\nu$ used here is different from the optical depth, $\tau (\lambda)$ (equation~(\ref{eq:tau})) presented in Section~\ref{sec:Non-mono}.

While computing the transmission spectra we neglect center-to-limb variation in the stellar disc and assume that the stellar disc emits a uniform specific intensity at a given frequency.  If $I_\star$ is the stellar specific intensity, then the specific intensity of the radiation for a given frequency after passing through the planetary atmosphere during transit is
\begin{equation}
  \frac{I_\nu}{I_\star} = e^{-\tau_\nu}.
\end{equation}
Therefore, the absorbed specific intensity will be 1-I$_\nu$/I$_\star$.

In our case, our plane of sky consists of 201 $\times$ 201 cells and we shoot 51 frequency/velocity dependent rays through each of the 201$^2$ grid elements. Since our grid is larger than the projected area of the stellar disc, we assign a zero specific intensity of the ray that is emitted from pixels outside of the stellar disc. Hence, integrating over all these rays and dividing by the flux of the star, we can calculate the transit depth as given below:
\begin{equation}
\Delta F_\nu = \frac{\int\int (1-e^{-\tau_\nu})dxdy}{\pi R_\star^2}
\end{equation}
where dx and dy are the cell sizes. Given the way we build our grids, we have dx = dy = 2 $\times$ 10R$_p$/201 $\sim 0.1\mathrm{R}_p$. In simple words, the transit depth $(\Delta F_\nu)$ is the amount of stellar flux absorbed by the planetary atmosphere which has different effective radius $(R_{\rm p}^{\rm eff})$ at different wavelengths and can be written as
\begin{equation}
\Delta F_\nu  = \frac{\pi(R_{\rm p}^{\rm eff})^2}{\pi R_\star^2}.
\end{equation}
The amount of stellar flux obscured by the planetary disc (without the atmosphere) for our hypothetical system is
\begin{equation}\label{eq:geom}
\Delta F_{\rm geom}  = \frac{\pi(R_{\rm p})^2}{\pi R_\star^2} = 0.021
\end{equation}
As hot Jupiters have hydrogen-dominated atmospheres,  the stellar hydrogen photons get absorbed in the planetary atmospheres and a cyclic variation in atmospheric properties leads to a variation of the absorbed stellar flux in Ly$\alpha$ and H$\alpha$ lines.

%For our exoplanetary system, we consider an impact parameter b = 0.5070 with a transit duration of 3.0648 h similar to HD209458

\begin{figure}
\includegraphics[width=0.52\textwidth]{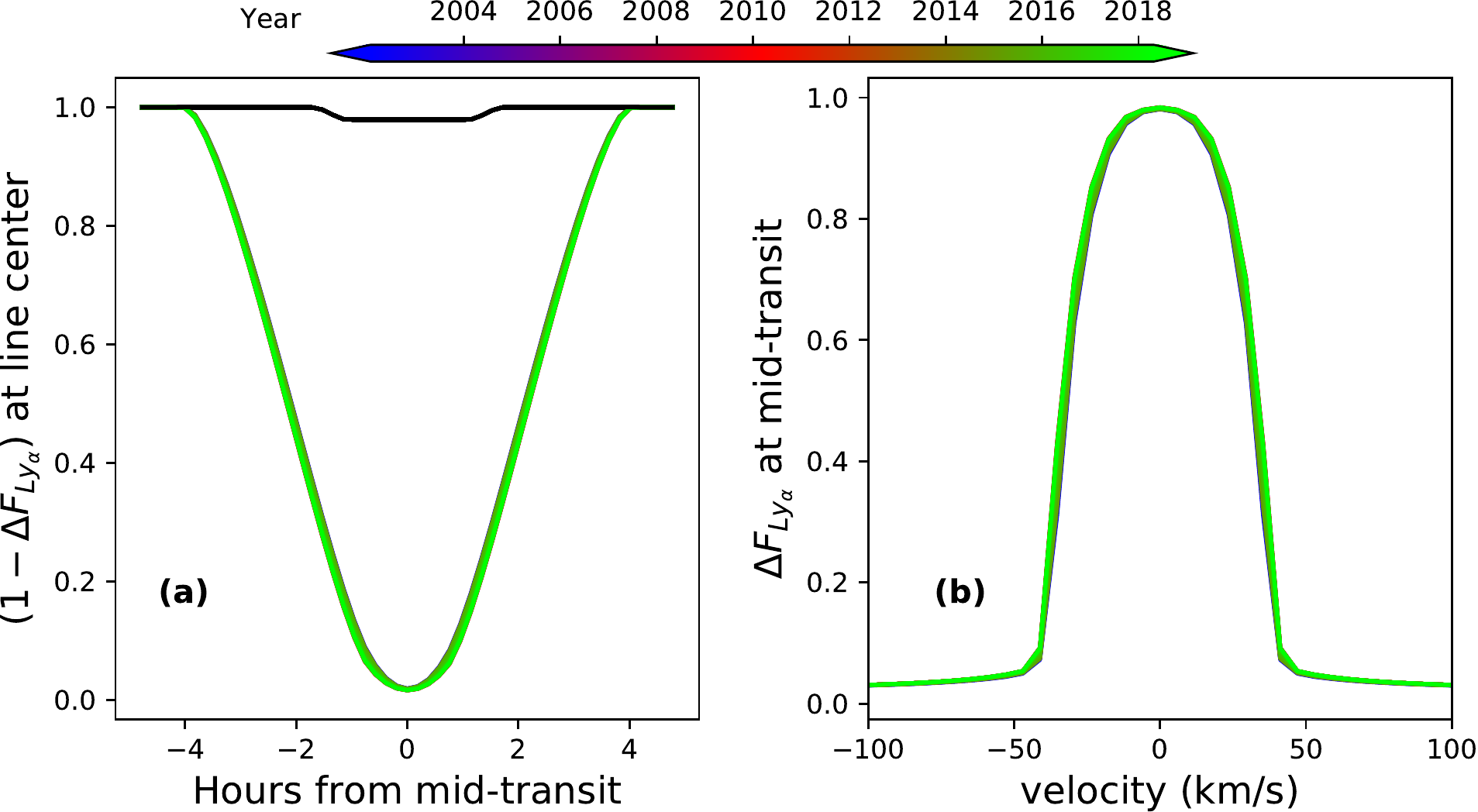}
\caption{(a) Ly$\alpha$ transit spectra at line center as a function of time from mid-transit. The black solid line shows the light curve for planet obscuration only. (b) Ly$\alpha$ line at mid-transit as a function of Doppler velocity. The colorbar represents the time span covering one and a half solar cycles.}
\label{fig:Lyalpha}
\end{figure}

\begin{figure*}
\includegraphics[width=0.95\textwidth]{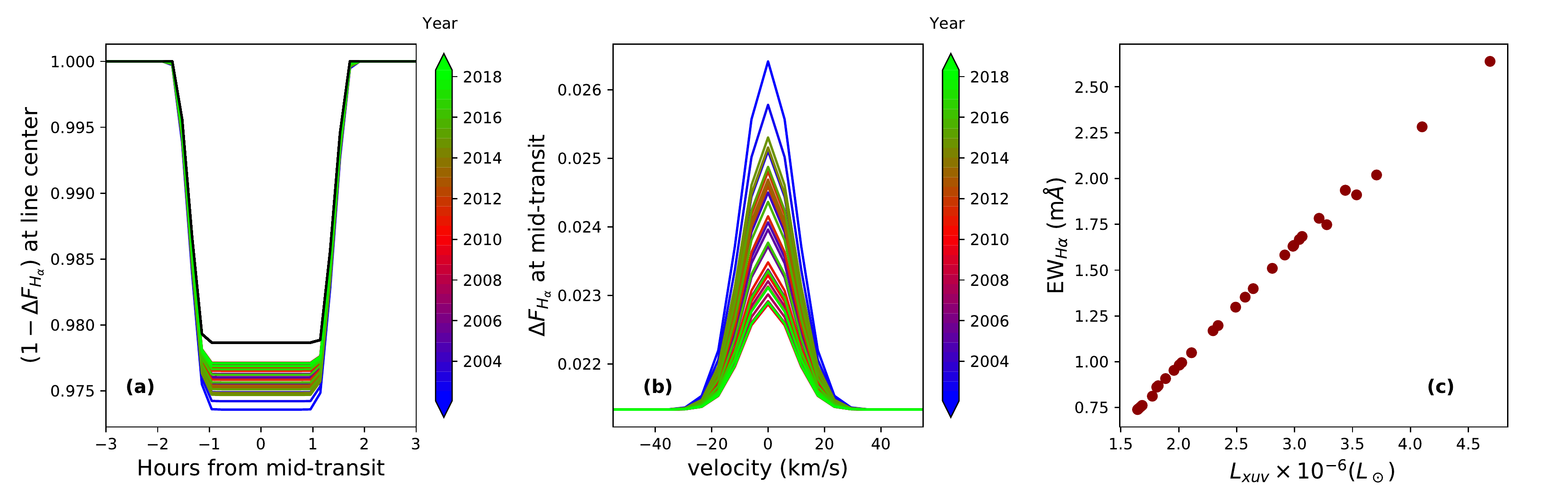}
\caption{(a) H$\alpha$ transit light curves as function of time from mid-transit, at line centre. The black solid line shows the geometric transit curve for planet obscuration only. (b) Transit depth in the H$\alpha$ line as a function of Doppler velocity at mid-transit. The colorbars represent the time covering one and a half magnetic cycles. (c) Variation of the equivalent width of the H$\alpha$ line versus the stellar L$_{\rm xuv}$ covering the same one and a half solar cycles. }
\label{fig:Halpha}
\end{figure*}

\subsubsection{Ly$\alpha$ transits}\label{sec:lyalpha}
The absorption spectrum in Ly$\alpha$ during planetary transit is shown in Fig.~\ref{fig:Lyalpha}.  Fig.~\ref{fig:Lyalpha}(a), show the synthetic transit light curves at the Ly$\alpha$ line centre over one and a half magnetic cycles of the Sun. Although the line center of Ly$\alpha$ is not accessible from observations, we plot 1-$\Delta F_{\rm{Ly}\alpha}$ at the line centre for an easy understanding of how much Ly$\alpha$ gets absorbed in the atmosphere of the hot Jupiter. For our exoplanetary system, we consider an impact parameter b = 0.5070 similar to HD209458b. The transit light curve for the geometric transit of the planet is shown in black solid curve. The absorption depth $\Delta F_{\rm geom}$ (see equation~(\ref{eq:geom})) due to the geometric transit of the planet is 0.021, indicating that most of the absorption in the Ly$\alpha$ line center in Fig.~\ref{fig:Lyalpha}(a) occurs due to an extended neutral hydrogen atmosphere around the hot Jupiter. Also it is evident from the peak value of the Ly$\alpha$ absorption at mid-transit that the neutral hydrogen density is high enough to absorb most of the Ly$\alpha$ flux from the host star.
Although we see a variation of a factor of 2.5 in escape rates, this variation is not significant to alter the Ly$\alpha$ transit profiles. This means that a Sun-like magnetic cycle would not change the Ly$\alpha$ transmission spectroscopic signature of a HD209458b-like planet. The colorbar in the Fig.~\ref{fig:Lyalpha}(a) shows the years which cover the one and a half solar cycles.

 In Fig.~\ref{fig:Lyalpha}(b) we show the absorption $\Delta F_{\rm{Ly}\alpha}$ in Ly$\alpha$ spectra at mid-transit as a function of Doppler velocity for all the 34 cases covering one and a half magnetic cycles. Here as well, we cannot observe any cyclic modulation in the Ly$\alpha$ spectra because of the same reason discussed in the previous paragraph.
 The symmetric profile in the blue and red wing is  due to the spherical symmetry considered in our atmospheric escape model.

\subsubsection{H$\alpha$ transits}
The absorbed stellar flux due to a transiting planetary atmosphere is very strong in the Ly$\alpha$ line center,  which is not observable. Unlike Ly$\alpha$, the H$\alpha$ line can be studied from the ground and is not affected by the interstellar absorption as Ly$\alpha$. In this section, we  calculate the transit spectra in the H$\alpha$ line. The temperature of the planetary atmosphere is sensitive to the cyclic changes of incident stellar radiation. This affects the population of hydrogen atoms in the first excited state n = 2. Thus, we expect to see changes in the absorption depth of H$\alpha$ over solar cycle.
Fig.~\ref{fig:Halpha} shows the H$\alpha$ transit spectra for different XUV fluxes over one and a half magnetic cycles. The absorption in H$\alpha$ at line center over the transit time similar to Fig.~\ref{fig:Lyalpha}(a) is shown in Fig.~\ref{fig:Halpha}(a). Contrary to the Ly$\alpha$ absorption, H$\alpha$ shows less absorption. The cyclic modulation of the H$\alpha$ absorption is small but noticeable. The maximum H$\alpha$ absorption of $\Delta F_\nu$ $\sim 2.64 \%$ is observed around year 2002, where L$_{\rm xuv}$ is maximum in our sample dataset. This absorption is 0.53$\%$ in excess of the geometric transit depth of 2.1$\%$. The minimum absorption of H$\alpha$ $\sim 2.29\%$ is observed during solar minima around years 2009 and 2018, where the L$_{\rm xuv}$ is minimum. Hence the amplitude in the H$\alpha$ transit depth during the cycle is 0.35$\%$. Our column densities for neutral hydrogen in the first excited state (n =2) range from 2.8 $\times$ 10$^{11}$ cm$^{-2}$ at minimum to 1.1 $\times$ 10$^{12}$ cm$^{-2}$ at maximum. The absorption in H$\alpha$ at mid-transit as a function of Doppler velocity over cycle is shown in Fig.~\ref{fig:Halpha}(b). We see that most of the absorption in H$\alpha$ occurs in the Doppler velocity range between $\pm 30$ km/s. It is likely that the absorption would extend to higher velocities if interactions with a stellar wind were considered \citep{Villarreal2018}. However, this is not captured in our 1D model.

To quantify the cyclic variation of the H$\alpha$ absorption, we calculate the equivalent width of the H$\alpha$ line as
\begin{equation}\label{eq:EW}
EW_{\rm H{\alpha}} = \int_{v_i}^{v_f} [\Delta F_\nu- \Delta F_{\rm geom}] dv
\end{equation}
where $v_i$ and $v_f$ are the extreme ranges of Doppler velocity over which we have performed the integration, and $\Delta F_{\rm geom}$ is the geometric transit depth as given by equation~(\ref{eq:geom}). In our simulations, we integrate the excess absorption in H$\alpha$ over the entire velocity range of $\pm500$ km/s, but note that nearly all the absorption occurs within $\pm 30$ km/s (see Fig.~\ref{fig:Halpha}(b)). In Fig.~\ref{fig:Halpha}(c), we plot the EW$_{\rm H{\alpha}}$ for our entire dataset over one and a half solar cycles as a function of L$_{\rm xuv}$. During the solar minimum the lowest L$_{\rm xuv}$ gives a EW$_{\rm H{\alpha}}$= 0.74 m$\Angstrom$ whereas during solar maximum, the estimated EW$_{\rm H{\alpha}}$ = 2.64 m$\Angstrom$. The reported upper limit of H$\alpha$ absorption observed in the HD209458 system (similar to our hypothetical system) is 1.7 m$\Angstrom$ \citep{Winn04}, which suggests that
a variation of 1.9m$\Angstrom$ between maximum and minimum of the cycle could be significant during transit measurements. If we use this upper limit to guide what could be potentially detectable, we infer that  H$\alpha$ transits could produce observable signatures around cycle maximum, due to an increase in column density of neutral hydrogen in the first excited state. We caution however that at cycle maximum, the background stellar H$\alpha$ is likely to be more `noisy', which requires extra care while interpreting transit observations.

\section{Magnetic field as a probe of XUV radiation for other stars}\label{sec:xuv}
The XUV radiation from the Sun is strongly correlated with its magnetic cycle. This
correlation is clear from the measurement of the XUV radiation \citep{Woods2005} as well as from the reconstructed XUV radiation \citep{Svalgaard2015}. Here we explore whether the relationship between solar magnetic field with its XUV flux can be extrapolated for others stars on their main sequence. Although resolving starspots in other stars is a difficult task, it is now possible to obtain the large-scale component of surface magnetic field in many Sun-like stars with the Zeeman Doppler Imaging (ZDI) technique \citep{Donati06, Fares13, Fares17}. With the mean magnetic flux for the other stars, we have the opportunity to derive the XUV flux of those stars by extrapolating the relationship between the magnetic flux and the XUV flux of the Sun.

\begin{figure*}
\includegraphics[width=\textwidth]{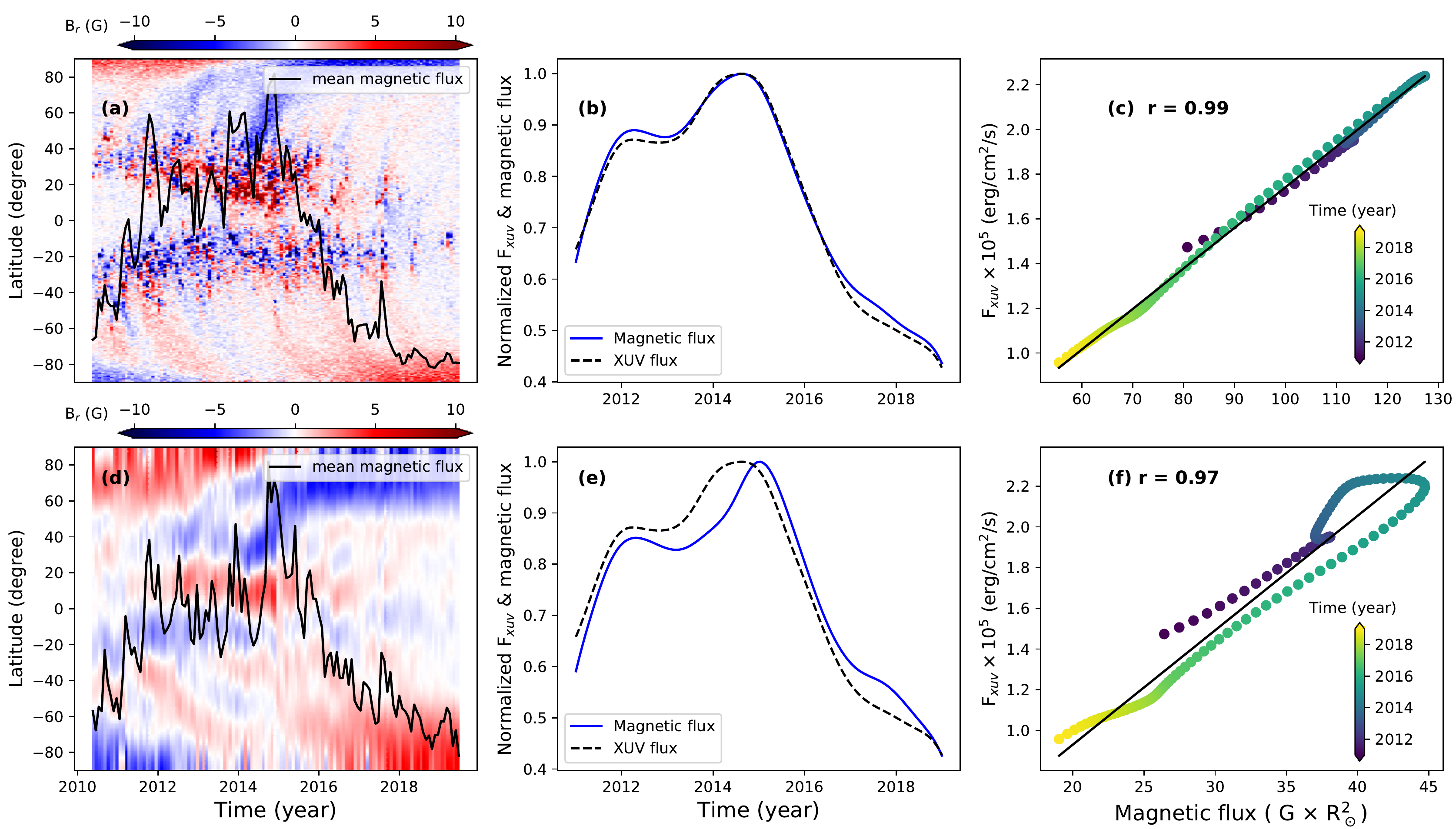}
\caption{Top row: (a) Butterfly diagram of the high resolution solar magnetic field is plotted for the solar cycle 24. This diagram is constructed from HMI synoptic maps, averaged over longitude. The solid black line represents the mean radial magnetic flux and ranges from $59$ to $146$ $G  R_\odot^2$. (b) Solar cyclic variation of normalized solar XUV flux and the surface magnetic XUV flux for the cycle 24. (c) Correlation between the solar XUV flux and surface magnetic flux over the cycle 24. The black solid line shows the best fit with a correlation coefficient $r= 0.99$. The colorbar represents the years covering the data for the cycle 24. Bottom row: (d), (e) and (f) are the same as (a), (b) and (c) respectively but only considering the large-scale component of the solar magnetic field up to a harmonic coefficient $\ell_{\rm max}$ = 10. The mean radial magnetic flux in (d) ranges from  $17$ to $58$ $G  R_\odot^2$. The correlation coefficient in (f) is $r = 0.97$, and shows a relationship on the form $F_{\rm xuv} \propto  \phi_b^{1.04 }$.}
\label{fig:mag_maps}
\end{figure*}

To derive this correlation, we first use the high resolution synoptic magnetic maps of the Sun for the solar cycle 24 starting from year 2010 to 2019 from the Helioseismic Magnetic Imager (HMI) on board of the Solar Dynamics Observatory \citep{Scherrer2012}. The azimuthally-averaged radial magnetic maps for the above mentioned period is shown in Fig.~\ref{fig:mag_maps}(a), in the format of a butterfly diagram. The mean radial magnetic flux integrated over each synoptic map is over plotted using black solid line. The direct comparison of the solar XUV flux with mean magnetic flux over solar cycle 24 is depicted in Fig.~ \ref{fig:mag_maps}(b). We have smoothed the mean magnetic flux data and XUV flux over a window of 1 year and normalized them for better comparison. The correlation plot between these two quantities is shown in Fig.~\ref{fig:mag_maps}(c). A very high correlation with a correlation coefficient of 0.99 is found between magnetic flux and XUV flux, which suggests that the surface magnetic flux of a star is a very good candidate for obtaining stellar XUV flux.

In Fig.~\ref{fig:mag_maps}(d) we plot the large-scale component of the surface magnetic field over cycle 24. We do this by filtering out the small-scale field from any given solar HMI synoptic map. In practice, we decompose the solar map using spherical harmonics as presented in \citet{2016MNRAS.459.1533V}. This allows us to calculate the spherical harmonics coefficients for each given harmonic order $\ell$. Note that the smaller scale structure is described by increasingly larger  values of $\ell$. Thus, to filter out the small-scale component, we only use the derived coefficients up to a maximum harmonic order of $\ell_{\rm max}$ = 10 for reconstructing the large-scale field. This method is commonly used in the literature to separate the large-scale field  \citep[e.g.,][]{2012ApJ...757...96D,2013ApJ...768..162P,Vidotto2018a,2018MNRAS.478.4390L}. Using only the large-scale component of the solar magnetic field is very important, because a given ZDI reconstructed map is restricted to the large-scale component of the stellar magnetic field. This is because ZDI suffers from magnetic flux cancellation, i.e., magnetic field of opposite polarities that fall within an element of resolution cancel out, allowing only the large-scale field to be reliably reconstructed \citep{2010MNRAS.404..101J,2011MNRAS.410.2472A,2014MNRAS.439.2122L}. Figures~\ref{fig:mag_maps}(e) and (f) show the mean large-scale magnetic flux and the XUV flux of the Sun as a function of time, and their correlation, respectively. The magnetic flux is calculated as
$$
\phi_b = \langle B_r\rangle 4 \pi R_\star^2
$$
where $\langle B_r\rangle$ is the average magnetic field of a synoptic map given in Gauss (G) and $R_\star$ is the solar/stellar radius. We see from Figs. \ref{fig:mag_maps}(e) and (f) that the solar XUV flux is  highly correlated also with the large-scale component of solar magnetic flux ($\ell_{\rm max}$ = 10 ), showing a correlation coefficient of 0.97. Note that the correlation decreases if we lower the value of $\ell_{\rm max}$ from 10 and it becomes very poor for $\ell_{\rm max} < 5 $. Fitting a power-law to this correlation, we find that
\begin{equation}
F_{\rm xuv} = 10^{3.642\pm 0.039}  \phi_b^{1.04 \pm 0.026},
\label{eq:powerlaw}
\end{equation}
where the XUV flux is given in erg cm$^{-2}$ s$^{-1}$ and the large-scale magnetic flux is given in $G R_\odot^2$.

This strong correlation between the large-scale component of the solar  magnetic field and its XUV radiation over a cycle has at least two possible explanations. In the simplest scenario, it could indicate that the large-scale field {\it directly} contributes to the solar XUV radiation. Alternatively, it could indicate that the large-scale field is correlated to another physical quantity that is responsible for the  XUV radiation. For example, it could be that the small-scale magnetic field, through a series of reconnection events, generates the high-energy solar irradiation and is thus correlated to \Fxuv . If there also exists a correlation between small-scale and large-scale fields, then it is natural to expect that the large-scale field is also correlated with \Fxuv . To the best of our knowledge, we are the first to report the existence of such strong correlation between solar surface magnetic flux (small and large-scales) and the solar XUV radiation (5--915\AA ). However, there are  observational studies that identified a strong correlation between X-ray and magnetic flux, going from small-scale structures in the Sun to large-scale stellar magnetism  \citep{Pevtsov2003, Vidotto2014}. In particular, \citet{Pevtsov2003} found a nearly linear correlation between the solar X-ray luminosity (2.8 -- 36.6\AA) and magnetic flux, which is similar to the one we report above. The correlation reported in  \citet{Vidotto2014} only considered the large-scale field of stars. However, contrary to the solar study in \citet{Pevtsov2003}, they reported a large spread in the X-ray -- magnetic flux correlation for stars. One possibility for their large spread is that magnetic fluxes and X-ray fluxes were not measured contemporaneously, and, in most cases, measurements were several years apart. Indeed, our solar study shows that, over cycle 24 (which was one of the weakest cycles in the Sun that we know of), both the XUV and magnetic fluxes varied by a factor of 3. Thus, using non-contemporaneous measurements naturally leads to increased spread in stellar correlations. A recent theoretical study by \citet{Blackman2015} explains the power-law dependence between X-ray luminosity and magnetic flux  based on dynamo magnetic field generation and magnetic buoyancy (these authors derived a power law exponent of $4/3$, see equation~(18) and footnote 2 in their paper).

%To make sure that our XUV correlation is not arising only from the X-ray part, we calculate the correlation of EUV part (125--915 \AA) of radiation with magnetic flux. We find a strong correlation of $r = 0.97$ between them.

\begin{figure}
\includegraphics[width=0.51\textwidth]{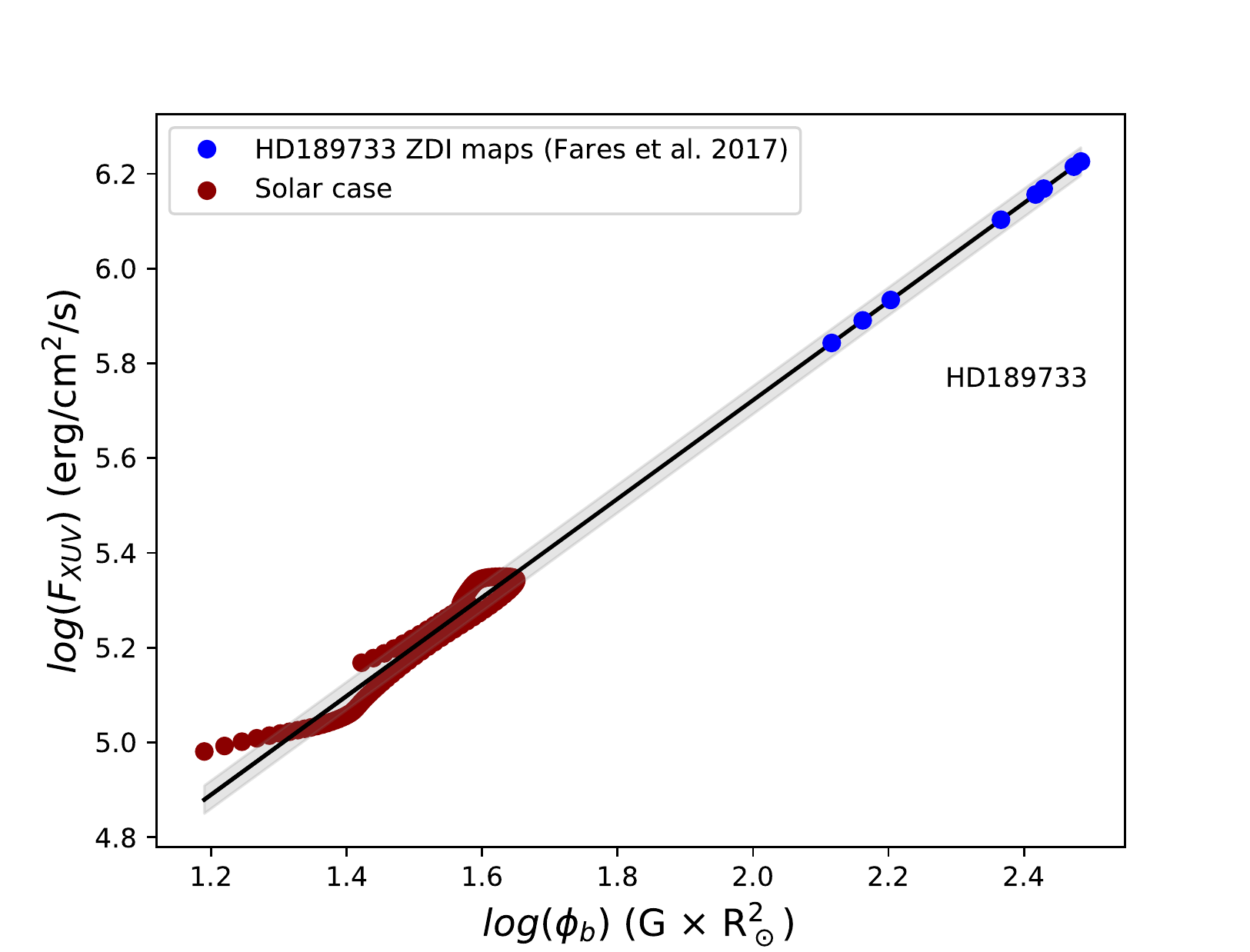}
\caption{Variation of solar XUV flux as a function of solar surface magnetic flux over the solar cycle 24 are shown using dark red filled circles. The black solid line shows the best fit with error bar in shaded regions (equation (\ref{eq:powerlaw})). The blue dots represent the XUV radiation extrapolated using the magnetic flux observed in the planet-hosting star HD189733 at 9 different epochs.}
\label{fig:xuv_corrl}
\end{figure}

The existence of a high correlation between solar surface magnetic flux with solar XUV flux motivates us to use surface magnetic maps of other stars as a proxy for their XUV flux. To do that, we apply the fit presented in equation (\ref{eq:powerlaw}) to stellar magnetic fluxes reported in ZDI observations.
Fig.~\ref{fig:xuv_corrl} shows the power law fit (black line) overplotted to the solar data (red points). The shaded area in the plot shows the error bar in the power law fit. The blue points are stellar magnetic fluxes derived from ZDI observations of the planet-hosting star HD189733.

HD189733 is a K2 star, whose magnetic field has been reconstructed using the ZDI technique at multiple epochs  \citep{2010MNRAS.406..409F, Fares13, Fares17}.
The magnetic field reconstructed at different times as given in \citet{Fares17} are shown in the 1st column of Table~\ref{tab:HD189733}. The corresponding magnetic flux is calculated at the same epochs adopting a stellar radius of $R_\star = 0.76 R_\odot$, and the stellar XUV fluxes are derived using equation~(\ref{eq:powerlaw}). The blue filled circles in Fig. ~\ref{fig:xuv_corrl} represent the computed \Fxuv\ on the extrapolated line and these values are tabulated in the 3rd column of Table~\ref{tab:HD189733}. We found that \Fxuv\ ranges from $7$ to $17 \times 10^5$ erg cm$^{-2}$ s$^{-1}$.

 We can also estimate the EUV radiation from HD189733 using the method presented in \citet{Chadney2015}, which derives the value of EUV flux from X-ray flux. Using the X-ray luminosity during pre-flare activity on year 2011 (see Table~1 of  \citet{Pillitteri11}), the relation in \citet{Chadney2015} (equation~15 of their paper) gives an EUV flux equal to 7.8 $\times$ 10$^5$ erg cm$^{-2}$ s$^{-1}$. Note that their estimate do not include the X-ray part of the spectrum. Nevertheless, to cross check if the EUV radiation obtained with our method gives similar values as those obtained from their method, we have considered the EUV radiation of the Sun in the wavelength range from 125 to 915 $\Angstrom$. We then recalculate the relationship between the high-energy flux (now only considering the EUV part) with the surface magnetic flux over a solar cycle and found that $F_{\rm EUV} \propto \phi_b^{0.95 \pm 0.024}$, which has a similar slope as equation~(\ref{eq:powerlaw}). Applying our EUV relation to the magnetic maps of HD189733, we find that during the maximum magnetic activity on 2013 September, the EUV radiation is F$_{\rm EUV}$ = 1.19 $\times$ 10$^6$ erg cm$^{-2}$ s$^{-1}$, and during the minimum activity on 2006 June, F$_{\rm EUV} = 5.32 \times 10^5$ erg cm$^{-2}$ s$^{-1}$. Our result is quite comparable with the estimated EUV flux from \citeauthor{Chadney2015}'s method. We will use our computed XUV fluxes to investigate atmospheric evaporation of HD189733b.

\begin{table}
	\centering
	\caption{Estimated \Fxuv\ for the planet-hosting star HD189733, computed at  epochs  (first column) when ZDI magnetic field observations are available. Second and third columns show the observed magnetic field and corresponding magnetic flux. Our derived values of \Fxuv\ are given in the 4th column. The last column shows our computed atmospheric escape rate of HD189733b, calculated using \Fxuv\ as an input.}
	\label{tab:HD189733}
	\begin{tabular}{lccccr} % four columns, alignment for each
		\hline
		Epoch & $B$ & $\phi_b$  & $F_{\rm xuv}$& $\dot{M}$ \\
		& (G) & ($G R_{\odot}^2$) & ($10^5$ erg cm$^{-2}$ s$^{-1}$) & ($10^{10}$ g s$^{-1}$) \\
		\hline
	June 2006 & 18 & 131 & 7.0  & 2.8  \\
    Aug 2006 & 20 & 145 & 7.8 & 3.1 \\
    June 2007 & 22 & 160 & 8.6 & 3.4 \\
    July 2008 & 36 & 261 & 14 & 5.5 \\
    June 2013 & 36 & 261 & 14 & 5.5 \\
    Aug 2013 & 41 & 298 & 16 & 6.3 \\
    Sep 2013 & 42 & 305 & 17 & 6.5 \\
    Sep 2014 & 32 & 232 & 13 & 4.6  \\
    July 2015 & 37 & 269 & 15 & 5.7 \\
    Strong Flare & -- & -- & 35 & 12 \\
		\hline
	\end{tabular}
\end{table}

\section{XUV radiation and its effect on the atmosphere of HD189733b}\label{sec:HD189733}
HD189733 hosts a hot Jupiter at 0.031 au, with a mass and radius of 1.142 M$_{\rm Jup}$ and 1.138 R$_{\rm Jup}$, respectively.
The host star of HD189733b is magnetically active with a mass of 0.82 M$_\odot$ and radius of 0.76 R$_\odot$.
As we mentioned earlier stellar radiation heats up the planetary atmosphere and gives rise to a transonic wind, thus evaporating the planetary atmosphere. The stellar radiation varies with the magnetic field of the host star and forces different escape rate over the time-scale of the magnetic field variability. We incorporate the derived stellar XUV fluxes from last Section into our 1D planetary wind model (Section~\ref{sec:Non-mono}) to calculate the evaporation rate as a function of time for this exoplanet.
Table~\ref{tab:HD189733} lists the planetary evaporation rate (last column).  Even though a cyclic behaviour has not been found for this star so far, we refer to a `maximum' and a `minimum'  magnetic activity to the observed epochs where the star exhibits overall the largest  and smallest magnetic fluxes, respectively. The ratio of calculated XUV radiation during maximum and minimum epochs of measured magnetic activity is 2.4. As a result, the escape rate over minimum and maximum epochs of the magnetic variability varies over a factor of 2.3, which is not too different from the solar case (Section~\ref{sec:cycle_outflow}). The terminal velocity also varies around 40 km/s with the magnetic activity but with a very small amplitude (less than $\pm 1$~km/s).

\begin{figure}
\includegraphics[width=0.45\textwidth]{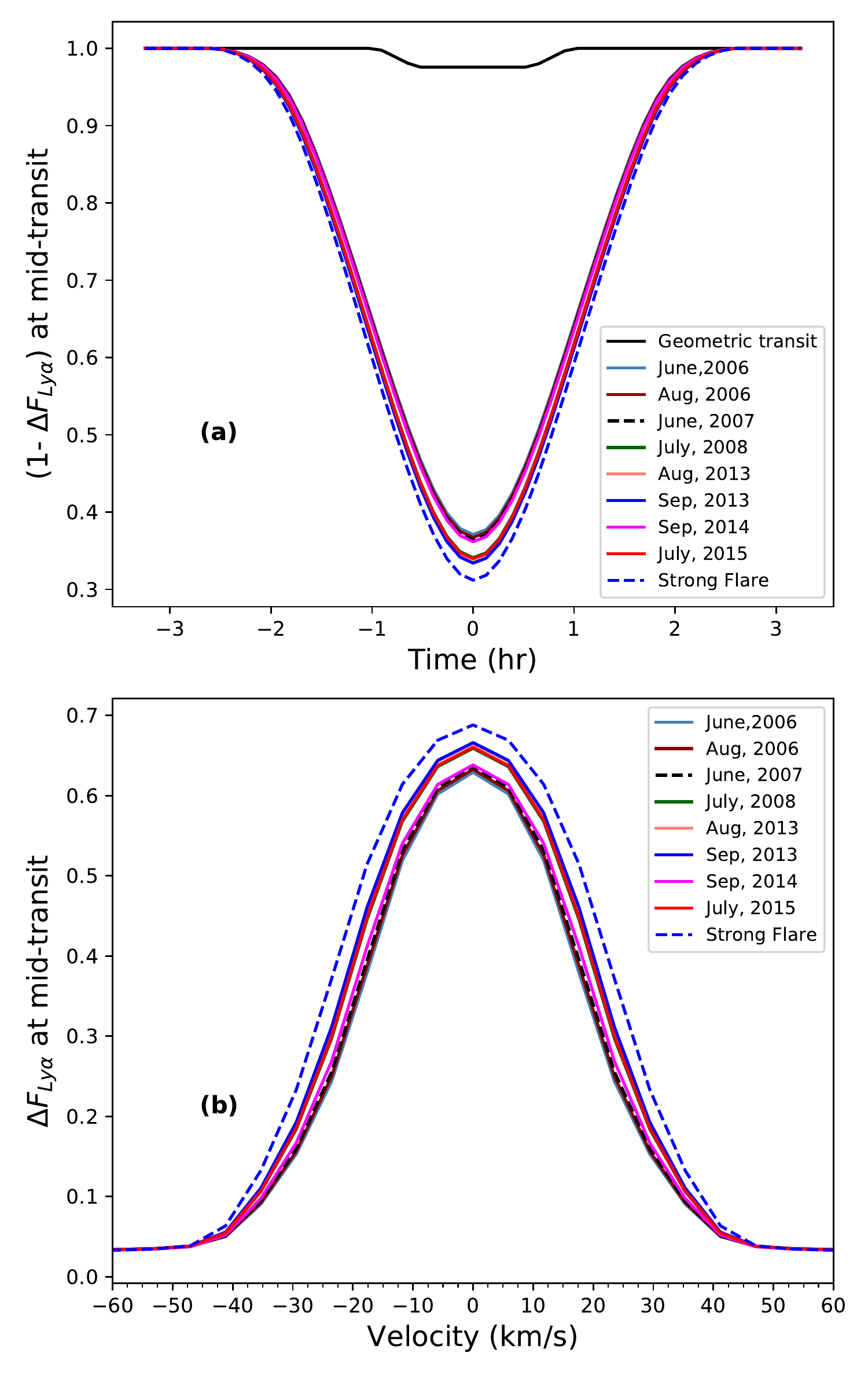}
\caption{(a) Transit light curve of HD189733b in Ly$\alpha$  at all epochs for which a ZDI magnetic map is available \citep{Fares17}. (b)  Transit depth of the Ly$\alpha$  line at mid-transit as a function of Doppler velocity for the same epochs. A hypothetical flare case, where \Fxuv\ is 5 times the value reported in June 2006, is shown by the  blue dashed lines.}
\label{fig:Lyalpha_hd189733}
\end{figure}

We also compute the spectroscopic transit spectra in the hydrogen lines for all epochs considered here following the same procedures as explained in Section~\ref{sec:transit}. The transit curves in the Ly$\alpha$ line are shown in Fig~\ref{fig:Lyalpha_hd189733}(a).
We find a significant transit depth ($\Delta F_{\rm Ly\alpha}$) in Ly$\alpha$, due to a strong evaporation of hydrogen. The solid black line in the figure represents the transit light curve due to the planet obscuration only, causing a depth at mid-transit of 2.4\%.
The simulated maximum transit depth during maximum magnetic activity (2013 Sept) and minimum activity (2006 June) at mid-transit is 67\% and 63\%, respectively, resulting in a  variation of about 4\% in transit depth over the studied epochs. Fig.~\ref{fig:Lyalpha_hd189733}(b) shows the transit depth as a function of Doppler velocity, where we see that most of the hydrogen absorption occurs within $\pm$ 50 km/s, and a temporal variation is observed in the blue and red wings for different epochs.

\begin{figure}
\includegraphics[width=0.45\textwidth]{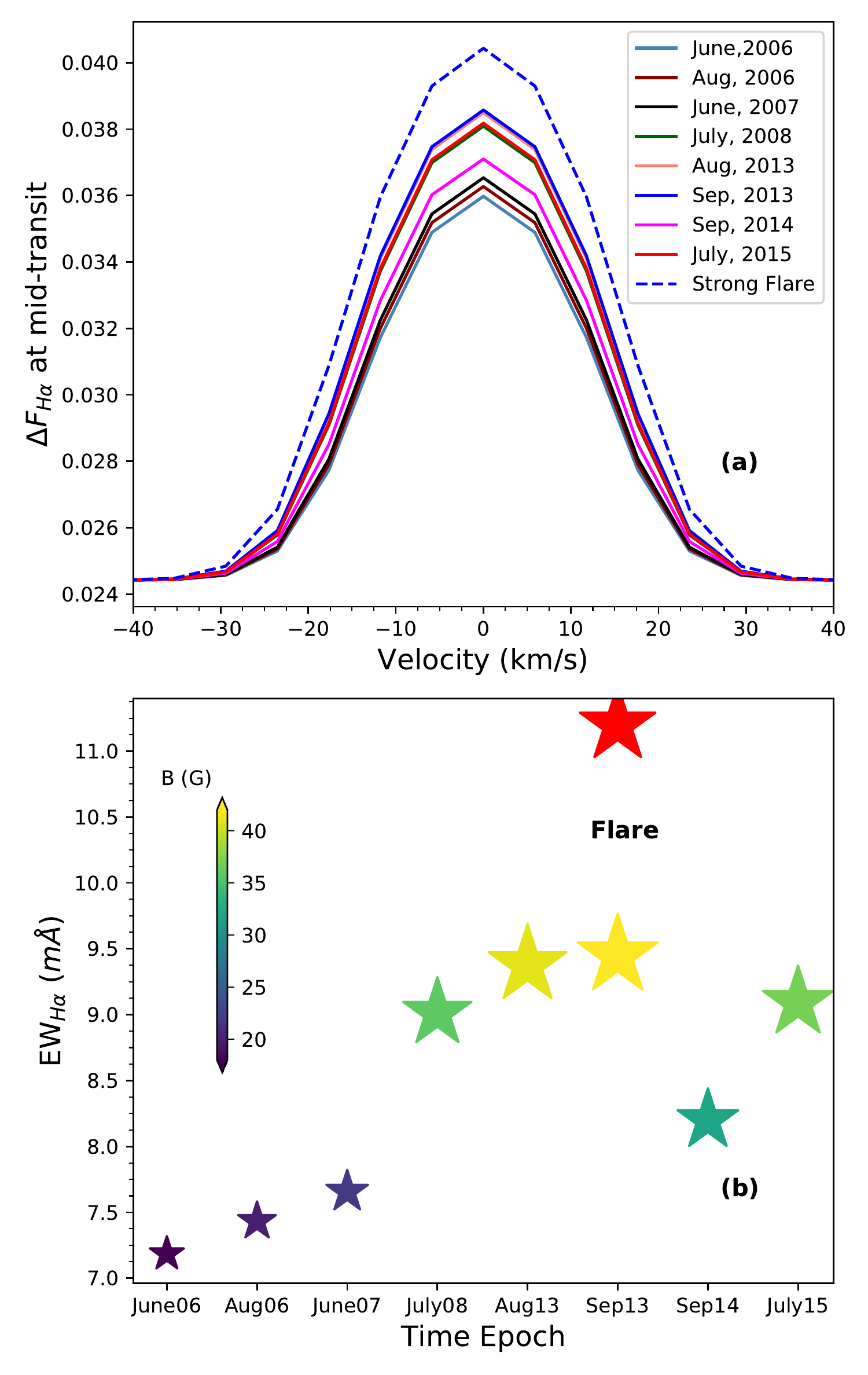}
\caption{(a) H$\alpha$ absorption spectra during mid-transit at different epochs. The flare case is shown by blue dashed line. (b) The equivalent widths (EW$_{\rm H{\alpha}}$) are plotted at the same epochs as (a). The colors in the EW$_{\rm H{\alpha}}$ plots are associated with the magnitudes of the magnetic field which are shown in the colorbar, except for the flare case, shown in red.}
\label{fig:Halpha_hd189733}
\end{figure}

A similar plot is shown for H$\alpha$ in Fig.~\ref{fig:Halpha_hd189733}(a). In this case, transit depths of 3.9\% and 3.6\% are found at maximum and minimum epochs, respectively. Given the geometric transit depth of 2.4\%, this implies  1.5\% and 1.2 \% excess in absorption during the maximum and minimum magnetic activity respectively. We show the H$\alpha$ equivalent width (following equation~(\ref{eq:EW})) in  Fig.~\ref{fig:Halpha_hd189733}(b) as a function of  the time of the observed magnetic field.
The size and the color of the individual point represents the strength of the magnetic field.
We see that EW$_{\rm H\alpha}$ varies with time quite significantly, from 7.2 m$\Angstrom$ at  minimum to 9.5 m$\Angstrom$ at maximum.

HD189733 is a relatively active star (at least for a planet host), which shows flaring activity. Interestingly, \citet{Lecavelier12} found an increase in atmospheric evaporation of HD189733b that took place 8h after an X-ray flare. The authors discuss whether it was an increase in the high-energy flux, caused by the flare, that enhanced atmosphere escape on the planet, or whether it was an associated coronal mass ejection that altered the stellar wind properties and thus affected  escape on the planet. Here, we simulate the former scenario. To emulate the effect of a flare on the escape of HD189733b, we assume an XUV flux that is five times stronger than the value at minimum. We name this case as a `strong flare' case, with a flux of $3.5 \times 10^6$~erg cm$^{-2}$ s$^{-1}$ or, equivalently, an XUV luminosity of $12.2 \times 10^{28}$ erg s$^{-1}$. Generally, we do not have a specific measurement of how the total XUV radiation  increases during flares. For the EUV band, \citet{Lecavelier12} reported an EUV luminosity of $7.1 \times$ 10$^{28}$ erg s$^{-1}$ at the epoch the flare was observed. This is only slightly lower than our flare case ($12.2 \times 10^{28}$ erg s$^{-1}$), but we note that our calculation is done over a wider wavelength band.

We run a planetary wind case using our calculated value of XUV flux for the `strong flare' case. We find that the escape rate increases by a factor of 4.4 over minimum activity, reaching a value of 1.2 $\times$ 10$^{11}$ g s$^{-1}$. The transit depth for this case is 69\% for Ly$\alpha$, which is 6\% deeper than the value we found at minimum. For H$\alpha$, the transit depth is  4\% (1.6\% excess over geometric transit), showing an equivalent width of 11.20 m$\Angstrom$. The flare results are shown by the blue dashed lines in Fig.~\ref{fig:Lyalpha_hd189733} and \ref{fig:Halpha_hd189733}(a) and by the red star in Fig.~\ref{fig:Halpha_hd189733}(b). The flaring state shows a significantly higher transit than all other cases considered for HD189733b.

It is interesting to compare our results with the observations from \citet{Lecavelier12} for Ly$\alpha$ and \citet{Barnes16} for H$\alpha$ transits. Note that our magnetic field data do not coincide with the epochs of these observations though. Regarding the Ly$\alpha$ transit, \citet{Lecavelier12} reported two  Ly$\alpha$ transits separated by 16 months. In the first set of observations, no evaporation was detected in the Ly$\alpha$ transit. This does not mean that the planet was not evaporating though. The Ly$\alpha$ line is absorbed at line centre and thus, it is possible that evaporation was taking place near the line centre and thus was undetected. In the second set of observations, which took place 8h after the flare, HD189733b presented a transit depth of 14.4\% in the blue wing of the Ly$\alpha$ line, indicative of escaping hydrogen. Although we found that the cyclic magnetic activity of the host star can introduce a significant temporal variation in the atmospheric properties of the planet, our results can not be directly compared to those from \citet{Lecavelier12}. This is because 1D models, like ours, do not consider the interaction with the stellar wind, which helps to accelerate the neutrals away from the star, broadening the transit line. The bulk of our absorption, for example, occurs within $\pm$ 50 km/s.

The comparison could be less problematic for  H$\alpha$. Our  H$\alpha$ transits show a variation of  0.27\% over the maximum and minimum activities of HD189733 and 0.45\% when considering the flare case. \citet{Barnes16} reported a temporal difference of 0.37\% in transit depths over two transits about one year apart. Their observed variation could, in principle, be caused by stellar cyclic-like variability. \citet{Barnes16} caution that their observed change could either be due to changes in the properties of planetary atmosphere alone, but it could also be artefacts introduced by stellar variability in the original stellar H$\alpha$ spectra.

\section{Conclusions}\label{sec:conclusion}
In this work, we investigated the effects of stellar cyclic magnetic activity on the escaping atmospheres of close-in giant planets. The stellar XUV flux is responsible for atmospheric escape, as it heats the atmosphere of planets and drive a bulk outflow. However, as we showed in this paper using solar data, the stellar XUV flux varies during magnetic cycles, being stronger (weaker) during maximum (minimum). Over the past one and a half solar cycles, the solar XUV has varied by a factor of 2.9. In previous solar cycles, this variation is expected to have been even stronger, as cycle 24 was not a particularly 'strong' cycle (XUV data are not available for older cycles).

To investigate how a stellar cycle can affect atmospheric escape in a typical hot Jupiter, we modelled atmospheric escape in a fictitious planet orbiting the Sun at 0.05 au. We chose planetary parameters that resemble that of HD209458b (mass, radius, orbital distance). In our models, we included  the  cyclic evolution of the solar XUV radiation as measured by the TIMED/SEE instrument \citep{Woods2005}. For our simulations, we chose 34 epochs evenly distributed over the past one and a half solar cycles.
Our atmospheric model allows for inclusion of the full spectral energy distribution (SED) of the host star, as well as for a monochromatic flux (i.e., at a single wavelength). The differences between these models are summarised in Table \ref{tab:nonmono}. We showed here that, although using the SED is more realistic, the monochromatic approximation is not a bad one. Given that deriving a SED on the XUV part of the spectrum for a star other than the Sun is challenging, we conducted our remaining calculations assuming monochromatic fluxes. To convert the solar SED into a `monochromatic' flux, we integrated the SED over wavelength and derived one XUV flux for each simulated epoch. We then used these 34 values as inputs for our atmospheric escape calculations.

We found that planetary escape follows a cyclic behaviour, with escape rates that are nearly proportional to the cyclic incident solar XUV flux. For the data considered here, the maximum XUV flux near solar maximum is 2.9 times stronger than at minimum, resulting in a planetary escape rate that is 2.5 times larger at maximum. The other basic atmospheric parameters, such as temperature, pressure and density, also present cyclic variations. A cyclic signature is also seen in the terminal velocity of the escaping atmosphere, although with a very small amplitude.

To investigate whether the cyclic variation in the atmospheric properties could introduce a variation in observable signatures, we also calculated  synthetic transit spectra in two hydrogen lines. Although we predicted a strong Ly$\alpha$ transit, we did not find significant changes in the line profile during the cycle. This indicates that Sun-like magnetic cycles might not cause observable changes in Ly$\alpha$ transits for HD209458b-like planets. However, a cyclic variation could be more easily seen in  H$\alpha$ transits, even though the H$\alpha$ transit signal is much weaker than that in  Ly$\alpha$. This is because the H$\alpha$ line is very sensitive to the atmospheric temperature of the planet that varies over cycle. The temperature alters the population of neutral hydrogen in the first excited state, thus affecting the strength of H$\alpha$ absorption.
For our hypothetical HD209458b-like planet, orbiting the Sun at close distances, we found that the H$\alpha$ transit depths vary from 2.3\% to 2.6\% during the solar cycle, with its equivalent width varying from 0.74 to 2.64 m\AA. Observations of H$\alpha$ transits of HD209458b reported a non-detection with upper limit of 1.7 m\AA\ in equivalent widths, which suggests that variations of 1.9 m\AA, between activity maximum and minimum, could potentially be detectable.
While the XUV flux and its cyclic evolution is now readily available for the Sun, this is not the case for other stars. Thus,
we presented here a method to infer stellar XUV flux from stellar magnetic field measurements.
Using simultaneous measurements of the solar magnetic field from the HMI instrument aboard SDO and solar XUV flux from the TIMED/SEE instrument for the cycle 24, we found a tight correlation between solar magnetic flux $\phi_b$ and XUV flux, $F_{\rm XUV}$. This correlation exists for either the high-resolution HMI synoptic magnetic map, as well as for the large-scale field of the Sun, and, for the latter, it can be described as
$$F_{\rm XUV} \propto \phi_b^{1.04\pm 0.026}.$$
We can then extrapolate this relation from the Sun to other Sun-like stars with magnetic field measurements to derive stellar XUV fluxes. More importantly, given that there are now planet-hosting stars that show magnetic field evolution (either in the form of a regular cycle or not), we can use stellar magnetic field data to infer how their XUV flux has varied over cyclic timescales and how this variation has affected atmospheric escape in exoplanets.

We applied our method to the planet-hosting star HD189733. Firstly, to derive the stellar XUV flux, we used the large-scale magnetic field values reported in Table 4 of \citet{Fares17} into our $\phi_b$ -- \Fxuv\ relation. From the nine available magnetic maps from mid-2006 until mid-2015, we found XUV fluxes ranging from 7 $\times$ 10$^5$ to 17 $\times$ 10$^5$ erg cm$^{-2}$ s$^{-1}$. Note that although the evolution seen in the magnetic field of HD189733 has not been recognised as a regular cycle so far, during these nine epochs, the star went through a phase of enhanced/reduced magnetic activity, which, for simplicity, we refer to as maximum and minimum.

Using these nine fluxes as input to our  atmospheric escape model for HD189733b, we then found a temporal variation in the escape rate ranging from 2.8 $\times$ 10$^{10}$ to 6.5 $\times$ 10$^{10}$ g s$^{-1}$. We found a small temporal variation in the Ly$\alpha$ transit depth with an amplitude of 4\% during the stellar `cycle' (going from 63 to 67\% over the cycle), which, like our solar-like cycle study, is unlikely to be distinguishable in the observations. For H$\alpha$, the transit depth shows an amplitude of 0.27\%  between minimum and maximum magnetic activity and corresponding changes in equivalent width of 2.3 m$\Angstrom$, from 7.2 m\AA\ at minimum to 9.5 m\AA\ at maximum. Hence stellar cycles, especially when stellar magnetic activity is correctly taken into account in transit observations, could present new opportunities for the observations of increased hydrogen escape.

As HD189733 is a magnetically active star with observed flares, we considered a separate `strong flare' case where the stellar XUV flux was assumed to be 5 times higher than its minimum XUV flux. The corresponding increase in escape rate for the strong flare case is a factor 4.4, reaching a value of $1.2 \times 10^{11}$~g s$^{-1}$.
The Ly$\alpha$ transit depth reaches 69\% (6\% deeper than at minimum), and the H$\alpha$ transit depth reaches 4\%, with an equivalent width of 11.2 m\AA.

Our calculations have shown that stellar cyclic magnetic activity affects the properties of atmospheric escape. Additionally, a very strong flare  will also have a significant effect on atmospheric escape, enhancing even more escape rates. Although the cyclic variation of the XUV flux due to the magnetic activity of the host star is incorporated in our model, we have not considered any magnetic field neither from the host star nor from the planet directly in our simulations. For the very close-in planets, magnetic fields can affect  atmospheric escape via magnetic reconnection \citep{Lanza13}. Also, the effect of stellar wind and its cyclic variation in the atmospheric escape is not included. Interaction of stellar winds with atmospheric outflow is known to affect the evolution of exoplanetary atmospheres \citep[e.g.,][]{McCann19,Vidotto2020,Villarreal2018}. The solar Ly-$\alpha$ line varies with cycle \citep{Lemaire2015, 2018ApJ...852..115K} and thus there is a possibility that radiation pressure on the planetary outflow, which we did not consider here, could affect mass loss. However, a recent study by \citet{Debrecht2020} showed that radiation pressure is not sufficiently strong to affect the planetary winds on hot Jupiters. Given that the XUV radiation drives the planetary outflow and that the modulation of planetary escape is sensitive to this variation, we conclude that considering the effects of the magnetic cycle on XUV radiation is an important step to characterise the effects of magnetic cycles on planetary atmospheres. In future works, we will explore the interaction of cyclic stellar winds \citep{2016MNRAS.459.1907N, 2019ApJ...876...44F} and cyclic magnetic fields \citep{2016MNRAS.459.4325M, 2018MNRAS.479.5266J, BoroSaikia18} with planetary outflows, using more sophisticated 3D modelling.

\section*{Acknowledgments}
We thank an anonymous referee for her/his useful comments. This project has received funding from the European Research Council (ERC) under the European Union’s Horizon 2020 research and innovation programme (grant agreement No 817540, ASTROFLOW). We acknowledge funding received from the ‘Irish Research Council Laureate Awards 2017/2018’. CVD acknowledges the funding from the Irish Research Council through the postdoctoral fellowship (Project ID: GOIPD/2018/659). CHIANTI is a collaborative project involving George Mason University, the University of Michigan (USA), University of Cambridge (UK) and NASA Goddard Space Flight Center (USA).

\section*{Data Availability}
The data underlying this article were accessed from TIMED/SEE Solar EUV Experiment [located at \url{http://lasp.colorado.edu/data/timed_see/level3/}]. The derived data generated in this research will be shared on reasonable request to the corresponding author.

%%%%%%%%%%%%%%%%%%%%%%%%%%%%%%%%%%%%%%%%%%%%%%%%%%



\bibliographystyle{mnras}
\bibliography{myref_exoplanet} % if your bibtex file is called example.bib

%\begin{thebibliography}
%\begin{thebibliography}{99}
%\end{thebibliography}

%%%%%%%%%%%%%%%%%%%%%%%%%%%%%%%%%%%%%%%%%%%%%%%%%%

%%%%%%%%%%%%%%%%% APPENDICES %%%%%%%%%%%%%%%%%%%%%

%\appendix


%%%%%%%%%%%%%%%%%%%%%%%%%%%%%%%%%%%%%%%%%%%%%%%%%%

% Don't change these lines
\bsp	% typesetting comment
\label{lastpage}
\end{document}